\documentclass[11pt]{article}

\usepackage[T1]{fontenc}
\usepackage[utf8]{inputenc}
\usepackage{lmodern}
\usepackage{microtype}          

\usepackage[margin=1in]{geometry}

\usepackage{amsmath}
\usepackage{amssymb}

\usepackage{graphicx}
\usepackage{float}
\usepackage{subcaption}
\usepackage[export]{adjustbox}  
\usepackage{booktabs}
\usepackage{tabularx}
\usepackage{tikz}

\usepackage{xcolor}
\usepackage{comment}

\usepackage{authblk}

\title{\textbf{Online Preplay and Offline Replay Optimize Action Selection
in a Closed-Loop Navigation Model}}

\author[ ]{Masud Ehsani}
\author[ ]{Sen Cheng}
\affil[ ]{Institut f\"ur Neuroinformatik, Ruhr-Universit\"at Bochum, Germany}
\affil[ ]{\normalfont\small
 {\texttt{Masud.Ehsani@ruhr-uni-bochum.de}},
{\texttt{sen.cheng@ruhr-uni-bochum.de}}}
\date{}

\usepackage[colorlinks=true, linkcolor=blue!60!black,
            citecolor=blue!60!black, urlcolor=blue!60!black]{hyperref}

\begin{document}

\maketitle
\begin{abstract}

Hippocampal neurons not only encode the animal’s current position, but also generate sequences representing potential future states and past locations.  During goal-directed navigation, the represented location sweeps ahead of the animal's location (online forward replay). After such tasks, during rest, neuronal sequences occurring during sharp-wave ripples replay previously explored trajectories (offline forward replay).
Following random foraging, however, neuronal sequences resemble Brownian motion, which does not correlate with behavior (diffusive replay). The mechanisms producing such diverse sequences, and their distinct effects  on spatial coding and navigation, remain unclear.
Here, we address these questions using a closed-loop model of navigation in a two-dimensional open-field environment based on a spiking neuronal network. In our framework, neuronal sequences are generated by a common mechanism driven by background noise and the balance of excitation and inhibition. Different replay modes arise solely from task-dependent synaptic modifications and serve different functions:  Online forward replay facilitates shortcut discovery, offline forward replay consolidates optimal sequences of actions,  and diffusive replay helps in generalizing the action and state map and homogenizing internal connectivity. 
Together, these results show that a single network architecture can flexibly generate various replay types and clarify how each contributes to optimizing goal-directed navigation behavior.

\end{abstract}

\section{Introduction}

The hippocampal formation supports navigation and memory through place cells, neurons that fire at specific environmental coordinates to encode an animal’s position \cite{OKEEFE1971171}. During active exploration, these cells organize into theta sweeps—rapid sequences nested within each 4–12 Hz theta cycle that "sweep" toward the animal's immediate future path \cite{foster2007hippocampal, johnson2007neural}. During periods of quiet wakefulness and sleep, these same cells are reactivated within high-frequency sharp-wave ripples (SWRs) \cite{buzsaki2015hippocampal}. Neural decoding \cite{zhang1998interpreting} reveals that this SWR-bound activity represents structured spatial sequences that manifest in distinct modes depending on behavioral context \cite{Diba, STELLA}.

Crucially, the recurrent CA3 network is thought to be the primary generator of these diverse sequences. Its extensive recurrent architecture, characterized by sparse but powerful excitatory connections, allows for the rapid retrieval of stored spatial associations, which are then paced by theta-frequency inhibition to form ordered sweeps during navigation \cite{lisman2009prediction, skaggs1996theta}. The same network also drives hippocampal activity during immobility; specifically, population burst activity in CA3 has been shown to be a prerequisite for the sequence replay observed in CA1 during sharp-wave ripples \cite{BurstsCA3sullivan2011relationships, nakashiba2009hippocampal}. These bursts are highly skewed in size \cite{buzsaki2015hippocampal} and typically last tens to a few hundred milliseconds, with statistics appearing log-normal or power-law depending on brain state and measurement scale \cite{priesemann2013neuronal}. Computational studies indicate that log-normal synaptic weight distributions in recurrent networks can produce log-normal bursts \cite{lognormalomura2015}, whereas power-law bursting emerges when networks operate near a critical phase transition \cite{beggs2003neuronal, scarpetta2023criticality, poil2012critical}. Together, these findings suggest that replay during sharp-wave ripples arises intrinsically from the dynamical regime of the CA3 recurrent network, highlighting the importance of a CA3 model in which population bursts serve as the primary mode of activity and the stored spatial map provides the replayed content.

To understand how such a map is physically instantiated, models of CA3 have historically framed the network either as a point attractor \cite{treves1994computational}, a continuous attractor \cite{colgin2010attractor, tsodyks1995associative, Azizi2013}, or a hybrid supporting discrete memories and continuous spatial maps \cite{rolls2024theory, rolls2018storage}. In continuous attractor networks, activity evolves on a manifold whose topology reflects the encoded variable, a principle extensively studied in grid cell networks both experimentally \cite{Gridgardner2022toroidal} and theoretically \cite{Gridguanella2007model}. However, whether CA3 implements a true two-dimensional plane attractor remains debated \cite{knierim2012attractor}. Experimental and computational evidence instead suggests that CA3 may operate as a high-dimensional recurrent system with metastable, quasi-continuous attractors \cite{sasaki2007metastability}, producing an energy landscape with shallow basins that enables memory recall and replay but does not sustain ideal continuous attractor dynamics. This metastability, together with tight excitation–inhibition balance, provides a potential mechanism for generating spontaneous population bursts capable of activating specific assemblies.

Beyond these internal dynamics, experimental studies investigate the specific contents and behavioral relevance of such sequence replay. In goal-directed tasks, the network produces "online" forward replay during pauses to depict future paths toward goals \cite{pfeiffer2013hippocampal}, as well as backward (or reverse) replay to evaluate recently traveled routes and propagate reward signals \cite{foster2006reverse, Diba}. During subsequent rest or sleep, the network continues to replay these paths to consolidate task-relevant memories and stabilize the internal map \cite{dupret2010reorganization, ji2007coordinated}. Conversely, after random foraging, the network generates diffusive replay, where sequences resemble random-walk dynamics rather than directed trajectories \cite{STELLA}. In parallel with these offline functions, the real-time "look-ahead" mechanism observed during theta sweeps is considered essential for maintaining spatial orientation and coordinating goal-directed movement \cite{lisman2009prediction}.

These considerations highlight a gap in the literature on hippocampal replay: existing models rarely combine (i) intrinsic generation of burst dynamics, (ii) spontaneous replay of spatially encoded content embedded in the network as metastable states, and (iii) the behavioral relevance of this replay on action selection. To fill this gap, we implement a closed-loop spatial navigation model composed of place cells and action-selection neurons \cite{brzosko2017sequential, ghazinouri, Ghazi2}. Place cells emerge from spatially tuned inputs within a recurrent excitatory and inhibitory network and project onto an action-selection ring attractor that determines the agent’s movements. Plasticity in both recurrent and place-to-action pathways allows the agent to learn optimal trajectories and reach reward locations.

We evaluate this system in two navigation tasks. In a goal-directed task analogous to the Morris watermaze \cite{morris1984developments, morris1982place}, online and offline forward replay activate the appropriate action cells for each place cell, enabling shortcut discovery and flexible strategy adjustment. In a random foraging task, diffusive replay produces local sequences resembling random walks \cite{STELLA}, which promote a more uniform spatial representation, strengthen recurrent connectivity, and generalize place-to-action mappings across the environment, ultimately improving subsequent goal-directed navigation. Altogether, these findings show that neuronal sequences, shaped jointly by learning and intrinsic network dynamics, can flexibly guide behavior by influencing action selection.

\section{Methods}
We use a modified version of the closed-loop spatial navigation model implemented in \cite{brzosko2017sequential,ghazinouri}, extending it to include a recurrent network capable of embedding and replaying sequences. Brzosko et al. \cite{brzosko2017sequential} introduced a simplified action selection network with dopaminergic modulation of the connections from place cells to action neurons via an eligibility trace, enabling delayed STDP-based synaptic changes \cite{gerstner2018eligibility}. This framework has been used to investigate place cell properties \cite{ghazinouri}, neuromodulation \cite{wert2022dopamine}, and the effects of STDP rules on reversal learning \cite{Ghazi2}.

\subsection{Model Architecture}

\begin{figure}[hbtp]
    \centering
    \includegraphics[width=1\linewidth]{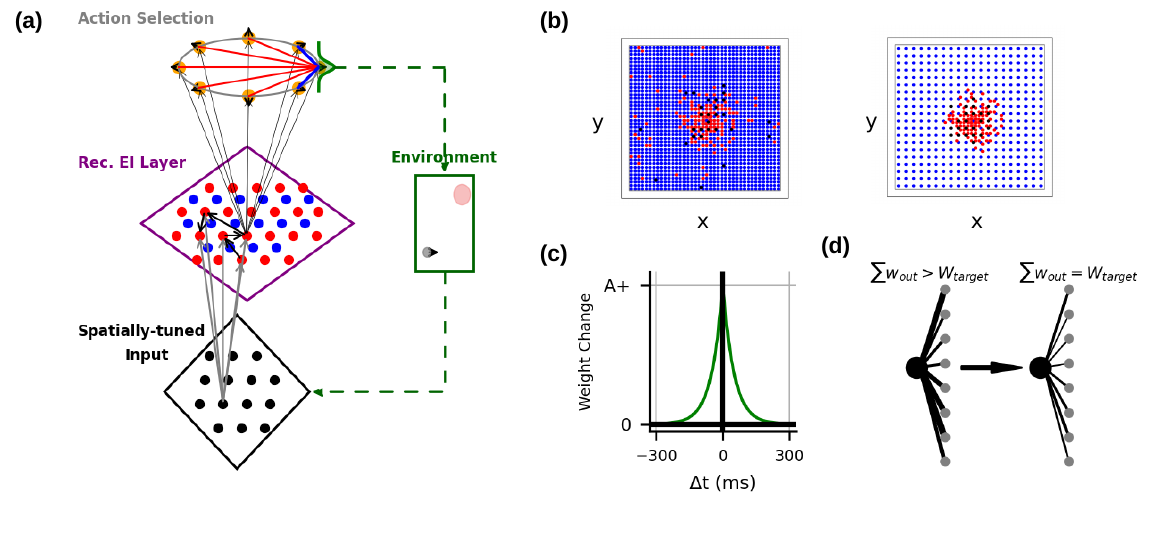}
    \caption{
            \textbf{Illustration of the network architecture and plasticity mechanisms.} 
\textbf{(a)} The network consists of three layers: the input, recurrent, and action-selection layer. The recurrent layer includes excitatory (blue dots) and inhibitory (red dots) populations. Only a subset of connections are indicated to illustrate the connectivity.
\textbf{(b)} Recurrent connectivity kernels. Excitatory (red) and inhibitory (black) neurons receiving connections from a central excitatory  (left plot)  and inhibitory (right plot) neuron.
\textbf{(c)} Symmetric STDP kernel
\textbf{(d)} Weight normalization scheme used for synaptic normalization. }
    \label{fig:net_scheme}
\end{figure}

Our modified model comprises three layers: a spatially tuned input layer received from the environment, a recurrent layer with place cells consisting of excitatory and inhibitory neurons, and an action-selection layer modeled as a ring attractor with effective Mexican hat connectivity, arising from local excitation and broader inhibition (Fig.~\ref{fig:net_scheme}a). The model operates in two distinct states: an online phase (exploration) and an offline phase (rest/replay). During exploration, the recurrent network receives spatially tuned input from the environment, and activity in the action-selection layer determines the agent's movement direction and velocity, computed as a population vector average (Appendix~\ref{subsec:ring_attractor}). In contrast, during the rest phase, the spatially tuned input is removed and replaced by spatially homogeneous excitatory and inhibitory noise modeled as Poisson spike trains.

Neurons in the recurrent and action-selection layers are modeled as leaky integrate-and-fire (LIF) units with conductance-based synapses. Excitatory and inhibitory synaptic inputs are implemented as transient conductance changes with distinct reversal potentials. During rest, network activity operates in a low firing-rate regime driven by spatially homogeneous external noise as well as internally generated fluctuations. An analytical approximation of the single-neuron gain function in this regime is provided in Appendix~\ref{subsec:Dynamics_sińgle_neuron_gain}. All neuron, synapse, connectivity, and simulation parameters are summarized in Appendix~\ref{subsec:param_table}.

The core of our model lies in the dynamics and plasticity of the recurrent layer. This layer receives environmental inputs, encodes trajectories to the reward location via its recurrent connections, and drives behavior through forward projections to the action-selection layer (the place-to-action pathway). Synapses in both the recurrent and place-to-action pathways employ dopaminergically modulated plasticity governed by eligibility traces.

The recurrent layer represents a spatial map of the environment as a 2D neural manifold. One can think of the cells as being arranged on this manifold such that neural assemblies encoding adjacent places lie closer to each other. These assemblies receive spatially tuned input from nearby locations and exhibit higher effective connectivity in both number and strength. More specifically, excitatory projections to both excitatory and inhibitory neurons follow a distance-dependent Gaussian connectivity kernel,
\begin{equation}
P(d) \propto \exp\left(-\frac{d^2}{2\sigma^2}\right),
\end{equation}
where $d$ denotes the distance between neurons on the manifold. The parameter $\sigma$ controls the spatial extent of local connectivity (Fig.~\ref{fig:net_scheme}b, left). In addition to this local structure, excitatory neurons form sparse long-range connections spanning the entire network. These connections have a fixed probability of 2\%, which is consistent with recurrent CA3 connectivity patterns \cite{CA3recurrent,CA3ishizuka1995}. Recurrent inhibition on this 2D network remains strictly local and follows a similar Gaussian kernel (Fig.~\ref{fig:net_scheme}b, right). To reflect distinct spatial footprints, we assume different values of $\sigma$ for excitatory and inhibitory populations. To prevent the agent from selecting invalid trajectories at the environment's edges, we
additionally include $N_{\text{border}} = 8$ border units (Table~\ref{tab:model_parameters}), each
associated with a distinct segment of the arena perimeter. A border unit acts as a parrot
relay, firing when the agent contacts or approaches its corresponding wall segment, and
projects a strong, static, inhibitory synapse ($w_{\text{border}} = -40$) onto the subset of
action-selection neurons whose preferred direction $\mathbf{a}_j$ points into that segment.
Because the ring attractor sustains a single localized bump through local excitation and
broader lateral inhibition (Appendix~\ref{sec:appendix}.2), suppressing the wall-directed sector of
action neurons destabilizes the bump locally; recurrent excitation from neighboring,
unobstructed headings then re-forms the bump in a permissible direction. This provides a
simple, non-plastic reflex -- the border synapses are static and are not subject to STDP or
synaptic normalization -- that automatically redirects the agent away from the boundary
without requiring an explicit, hard-coded turning rule.

While the model assumes a continuous manifold, this connectivity supports metastable rather than stable attractor states. By embedding this manifold within a larger balanced network, the system operates near a phase transition or critical regime. By supporting metastable states, the network avoids becoming trapped in a single activity pattern and instead transitions fluidly between neural assemblies through neuronal avalanches—spontaneous bursts of activity following power-law statistics. This regime provides a functional advantage: balanced inhibition suppresses runaway excitation while amplifying local fluctuations required for the reactivation and consolidation of structured sequences. The emergence of near-critical dynamics and neuronal avalanches in this regime is characterized quantitatively in Results Section~3.1.

\subsection{Learning Rules: Dopaminergic STDP and Synaptic normalization }

We use a variant of STDP with an eligibility trace \cite{gerstner2018eligibility},
gated by a neuromodulatory reward signal, so that only synapses active along the
agent's path to the goal during exploration are potentiated:
\begin{equation}
\begin{aligned}
\dot{w} &= c\,(n - b), \\
\dot{c} &= -\frac{c}{\tau_c} + C_1\,\Delta_{\text{STDP}}(t), \\
\dot{n} &= -\frac{n}{\tau_n} + \frac{C_2}{\tau_n}\,\delta(t - t_n).
\end{aligned}
\label{eq:eligibility}
\end{equation}
Here $c$ is the eligibility trace, $n$ the neuromodulator concentration, and $b$ a
baseline concentration acting as a threshold, so that weights grow only when $n > b$;
$t_n$ is the time of a neuromodulatory spike, $\tau_c$ and $\tau_n$ are the decay time
constants of $c$ and $n$, and $C_1$, $C_2$ are scaling constants.

Dopamine prolongs the plasticity window and converts the typically asymmetric STDP
kernel into a symmetric one \cite{brzosko2015retroactive}, and recurrent CA3 synapses
show a broad, symmetric, purely potentiating STDP curve that favours bidirectional
connectivity between co-active neurons \cite{mishra2016symmetric}. We therefore define
$\Delta_{\text{STDP}}$ through pre- and post-synaptic traces with all-to-all spike pairing. For a single spike pair with $\Delta t = t_{\text{post}} - t_{\text{pre}}$ this gives
\begin{equation}
\Delta c(\Delta t) = C_1\, e^{-|\Delta t|/\tau_{\text{STDP}}},
\label{eq:sym_kernel}
\end{equation}
potentiation for either spike order with no depressive lobe
(Fig.~\ref{fig:net_scheme}c). We set $\tau_{\text{STDP}} = 71\,\mathrm{ms}/\ln 2
\approx 102\,\mathrm{ms}$, so that the kernel matches the 142\,ms half-duration measured
at CA3--CA3 synapses \cite{mishra2016symmetric}, and $C_1 = 0.002$ for
excitatory--excitatory and excitatory--inhibitory connections
(Table~\ref{tab:model_parameters}).

During rest there is no explicit reward signal; instead, a constant tonic
neuromodulator level $n_0 > b$ is assumed. In this permissive regime, learning reduces
to unsupervised symmetric STDP.

To counteract the destabilizing positive feedback of purely potentiating STDP, we also
make excitatory-to-inhibitory (Exc--Inh) connections plastic, using the same symmetric
kernel (Eq.~\ref{eq:sym_kernel}, $\tau_{\text{STDP}} \approx 102\,\mathrm{ms}$). Because
potentiation of Exc--Inh synapses strengthens feedback inhibition onto co-active
assemblies, this acts on short timescales as a balancing mechanism that limits the
spread of activity \cite{kullmann2012plasticity,inhibition2008long}. Throughout this
study, plasticity is restricted to synapses originating from excitatory neurons. Furthermore, we implement a multiplicative synaptic normalization process during rest periods to maintain homeostatic firing rates \cite{turrigiano2008self}. After plasticity updates, the synaptic weight $w_{ij}$ from presynaptic neuron $j$ to postsynaptic neuron $i$ is rescaled to keep the total output weight from neuron $j$ constant at $w_{\text{target}}$ (Fig.~\ref{fig:net_scheme}d):

\begin{equation}
w_{ij} \leftarrow \frac{w_{ij}}{\sum_k w_{kj}} \cdot w_{\text{target}}
\end{equation}

\section{Results}

\subsection{Spatially Balanced E--I Network at the Edge of Instability Reproduces Burst Dynamics}
\label{sec:Balanced_net}

We reproduce the main characteristics of place cell activity during active exploration and rest within our low-dimensional network.
Specifically, this includes the emergence of a localized activity bump in response to spatial input at a given location, as well as population bursts of varying sizes during rest \cite{buzsaki2015hippocampal}. Although CA3 bursts tend to coincide with Down–Up transitions in the dentate gyrus \cite{Burstskajikawa2022up}, intracellular recordings show that many CA3 pyramidal neurons become hyperpolarized or show weak modulation during these transitions, consistent with strong recurrent or feedforward inhibition within the CA3 network \cite{hahn2007differential}. This suggests that DG-driven input can transiently excite subsets of CA3 neurons at Up-state onset, but that during sustained Up states or replay, CA3 operates under powerful inhibitory control, resulting in sparse, state-dependent firing \cite{Burstskajikawa2022up,hahn2007differential}. Thus, experimental evidence suggests that the network operates in a slightly inhibition-dominated regime.

In our model, we tune the external homogeneous inputs so that interneurons are slightly more depolarized than pyramidal cells, thereby maintaining inhibitory control. In Appendix ~\ref{subsec:Dynamics_sińgle_neuron_gain}, we analyze the firing of neurons receiving balanced subthreshold input, identify the appropriate balanced input rate that leads to low background firing of around 0.1 Hz, and approximate a low-firing regime gain function. Furthermore, for individual neurons to be able to trigger activity through recurrent connections, the synaptic weights should be relatively large, while for the system to support low-firing solutions, the average excitation and inhibition should approximately cancel out \cite{van1996chaos,brunel2000dynamics}. To achieve this balance, while maintaining an adequate level of fluctuations, we scale the weights as $w \sim 1/\sqrt{k}$, where $k$ is the number of connections per neuron.  

\begin{figure}[hbtp]
    \centering
    \includegraphics[width=1\linewidth]{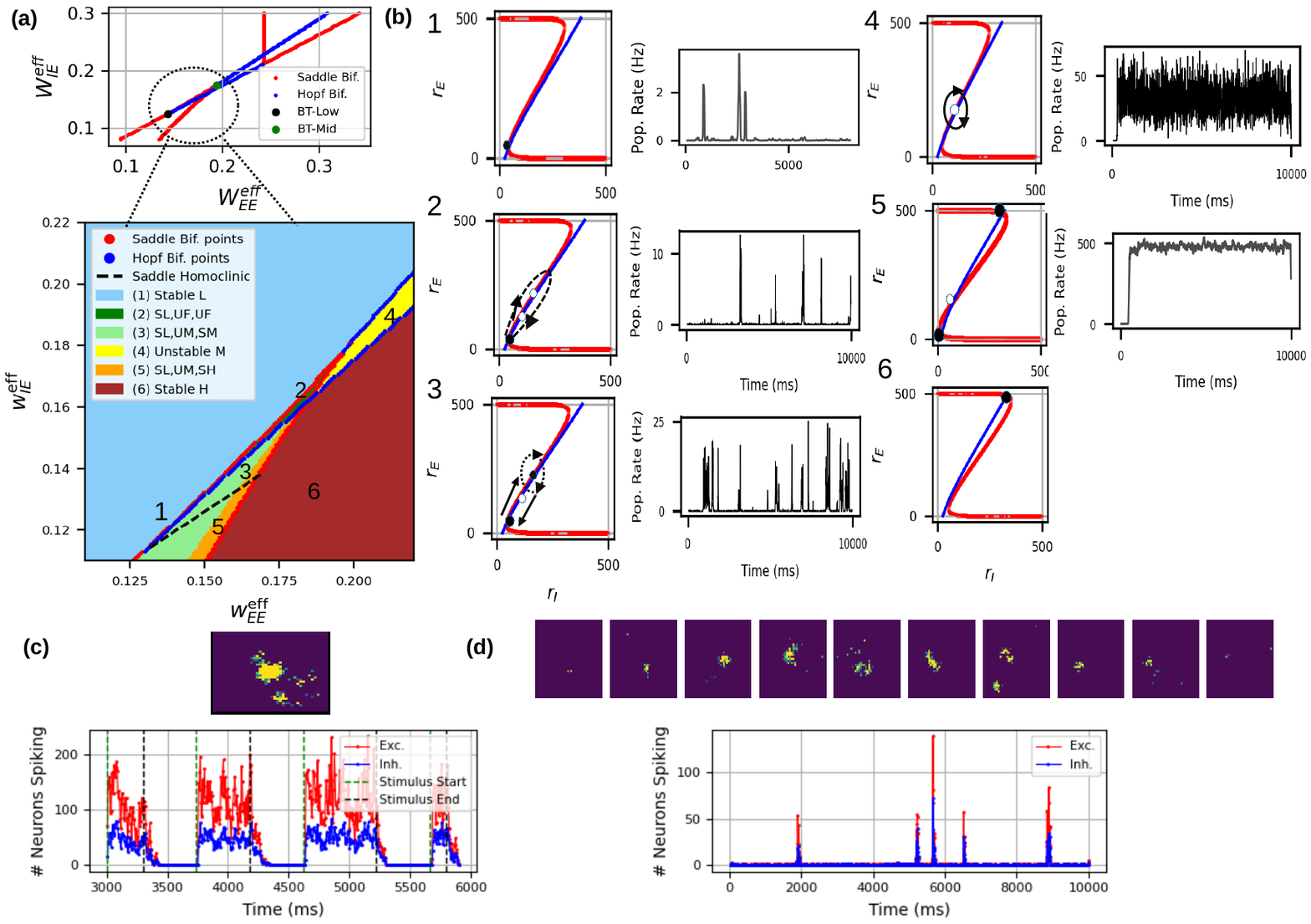}
        \caption{\textbf{Bifurcation diagram and network activity across the operating parameter regime.}
    \textbf{(a)}  Homogeneous bifurcation ($k=0$) diagram of the network dynamics in the plane of $
W_{IE}^{\mathrm{eff}} = W_{IE} \left(V_R^E - V_{\mathrm{th}}\right)$ and $ 
W_{EE}^{\mathrm{eff}} = W_{EE} \left(V_R^E - V_{\mathrm{th}}\right)$, where $V_R^E$ denotes the excitatory synaptic reversal potential. Saddle-node bifurcations on the low, middle, and high firing branches of the nullclines are shown as blue lines, the Hopf bifurcation as a red line, and the Bogdanov–Takens point (at the intersection of the saddle-node and Hopf bifurcation lines) is marked with black dots. The dashed line indicates a saddle homoclinic bifurcation. 
    \textbf{(b)}  Excitatory and inhibitory nullclines, along with population excitatory rates, are shown for the regions indicated on the bifurcation diagram.
    \textbf{(c)} Network activity in response to a pulse input applied to neurons at the center.  
    \textbf{(d)} Spontaneous activity driven by uncorrelated noise.  }
    \label{fig:bif_diag}
\end{figure}

This spatially balanced network can exhibit multiple dynamic states depending on the input and recurrent weight kernels. In Appendix~\ref{subsubsec:Bifurcation Analysis}, we derive the conditions on the recurrent weights for the system to operate near a critical phase transition. This regime gives rise to bursts with highly variable sizes and durations as well as irregular inter-burst intervals. By analyzing the bifurcations of the homogeneous mode ($k=0$), we find that the spatially connected network displays scale-free avalanches near the Bogdanov–Takens (BT) bifurcation point (Fig.~\ref{fig:bif_diag}a). Similar behavior has been reported in sparse and all-to-all homogeneous networks \cite{benayoun2010avalanches,ehsani2023scale}. We focus on the network operating near the Hopf bifurcation line specifically in the vicinity of the Bogdanov–Takens point. At the edge of this bifurcation, unstable fixed points characterized by eigenvalues close to zero exist within the low-firing regime (Fig.~\ref{fig:bif_diag}b2). Recurrent connections amplify fluctuations in external input to trigger neuronal avalanches. These events maintain a tight balance between inhibition and excitation. This precise balance allows activity to propagate with highly variable avalanche sizes and durations, following power-law distributions. Through a subcritical Hopf bifurcation, a low-firing fixed point surrounded by an unstable limit cycle emerges (Fig.~\ref{fig:bif_diag}b3). At the boundary between regimes 2 and 3 in the bifurcation map, stochastic fluctuations drive the system to switch intermittently between the quiescent and low-firing states. It is in this transition region that neuronal avalanches emerge, as recurrent interactions amplify noise-driven excursions while preserving overall excitation–inhibition balance. 

\begin{figure}[hbtp]
    \centering
    \includegraphics[width=1\linewidth]{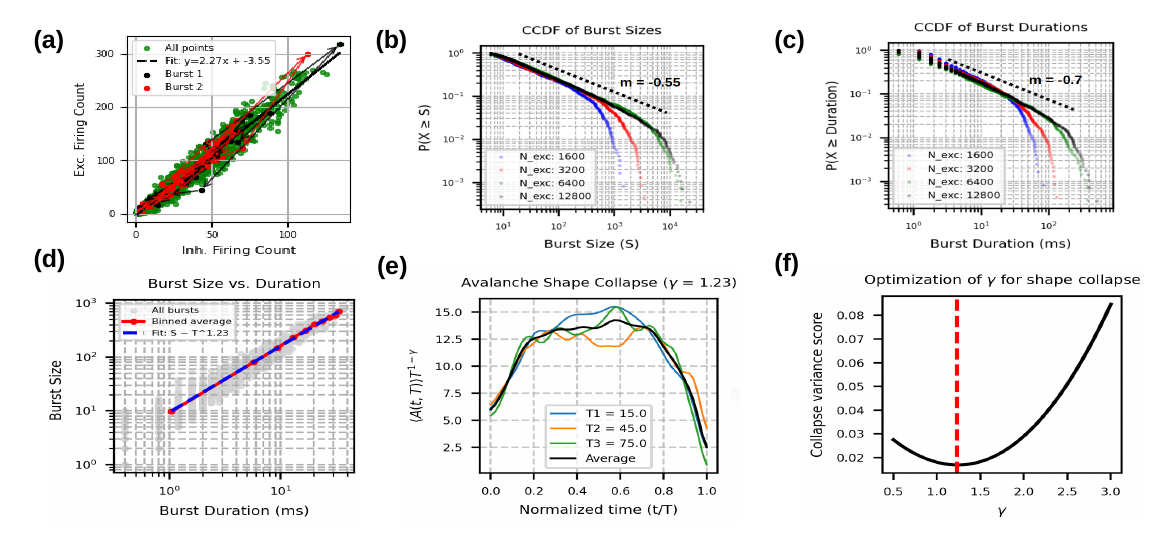}
    \caption{
    \textbf{Burst characteristics when the network is tuned to the edge of instability, near the Bogdanov–Takens bifurcation point (System No.1 in the bifurcation map)} 
    \textbf{(a)} Instantaneous excitatory versus inhibitory activity, binned in 5 ms intervals;  
    \textbf{(b,c)} Distribution of burst sizes (b) and  
    burst durations (c).
    \textbf{(d)} Average size vs. average burst duration.
    \textbf{(e)} Temporal profiles 
    of three bursts with different durations ($T$) are shown. The time axis is 
    normalized by the total duration ($t/T$), and the amplitude is scaled by 
    $T^{1-\gamma}$. The collapse onto a single universal curve indicates 
    self-similar dynamics across scales.
    \textbf{(f)} The best-fit exponent $\gamma$ for the shape collapse was determined by minimizing the variance score (mean squared residual) across the rescaled temporal profiles.
    }
    \label{fig:avalanche_char}
\end{figure}

In the same network, when spatially tuned input is applied, strong local inhibition gives rise to fast ($\approx50$~Hz) oscillations (Fig.~\ref{fig:bif_diag}c). Under conditions of near-balanced excitation and inhibition, network activity instead decays slowly following stimulus offset. During spontaneous activity in the presence of homogeneous background noise, we observe outward-propagating activity resembling local avalanches or population bursts (Fig.~\ref{fig:bif_diag}d). During these bursts, excitatory and inhibitory population rates satisfy an approximately linear relationship  (Fig.~\ref{fig:avalanche_char}a), suggesting that inhibition shadows excitation with a roughly constant average proportionality. To characterize avalanche statistics, we fit power-law distributions to the avalanche size and duration. 
The fits are performed on the complementary cumulative distribution functions (CCDFs) shown in 
Fig.~\ref{fig:avalanche_char}b,c. 
The corresponding probability density functions follow power laws of the form 
$p(S) \propto S^{-\tau}$ and $p(T) \propto T^{-\alpha}$. 
For a power-law--distributed variable, the CCDF exponent is smaller by one than the corresponding 
PDF exponent. 
Accordingly, the resulting power-law exponents for the size and duration PDFs are 
$\tau = 1.55$ and $\alpha = 1.7$, respectively.

The distribution of avalanche size shows evidence for a power-law (likelihood ratio test \cite{clauset2009power} against log-normal distribution, $p\approx 0.04$ across multiple runs).
In addition, the average burst size versus duration is well described by a power function with exponent $\gamma = 1.23$ (Fig.~\ref{fig:avalanche_char}d). The scaling relation
\[
\gamma = \frac{\alpha - 1}{\tau - 1}
\]
is also satisfied, with only small deviations attributable to finite-size effects. Furthermore, the shape collapse of avalanches indicates scale-free network activity near Hopf bifurcation line in vicinity of the BT bifurcation point (Fig.~\ref{fig:avalanche_char}e \& f). 
The power-law exponent values are in good agreement with experimental findings on spontaneous activity in anesthetized rats \cite{fontenele2019criticality}, supporting the conclusion that criticality emerges at the boundary between asynchronous and synchronized cortical states \cite{ poil2012critical,dalla2019modeling} . Moreover, the power-law exponents depend on network dimensionality and architecture. In Appendix~\ref{subsubsec:StochasticDynamics}, we analyze the scale-free avalanche regime in 2D networks with local excitation and in random sparse networks along the Hopf bifurcation line. In both network types, scaling and avalanche shape collapse hold.

\begin{figure}[hbtp]
    \centering
    \includegraphics[width=1\linewidth]{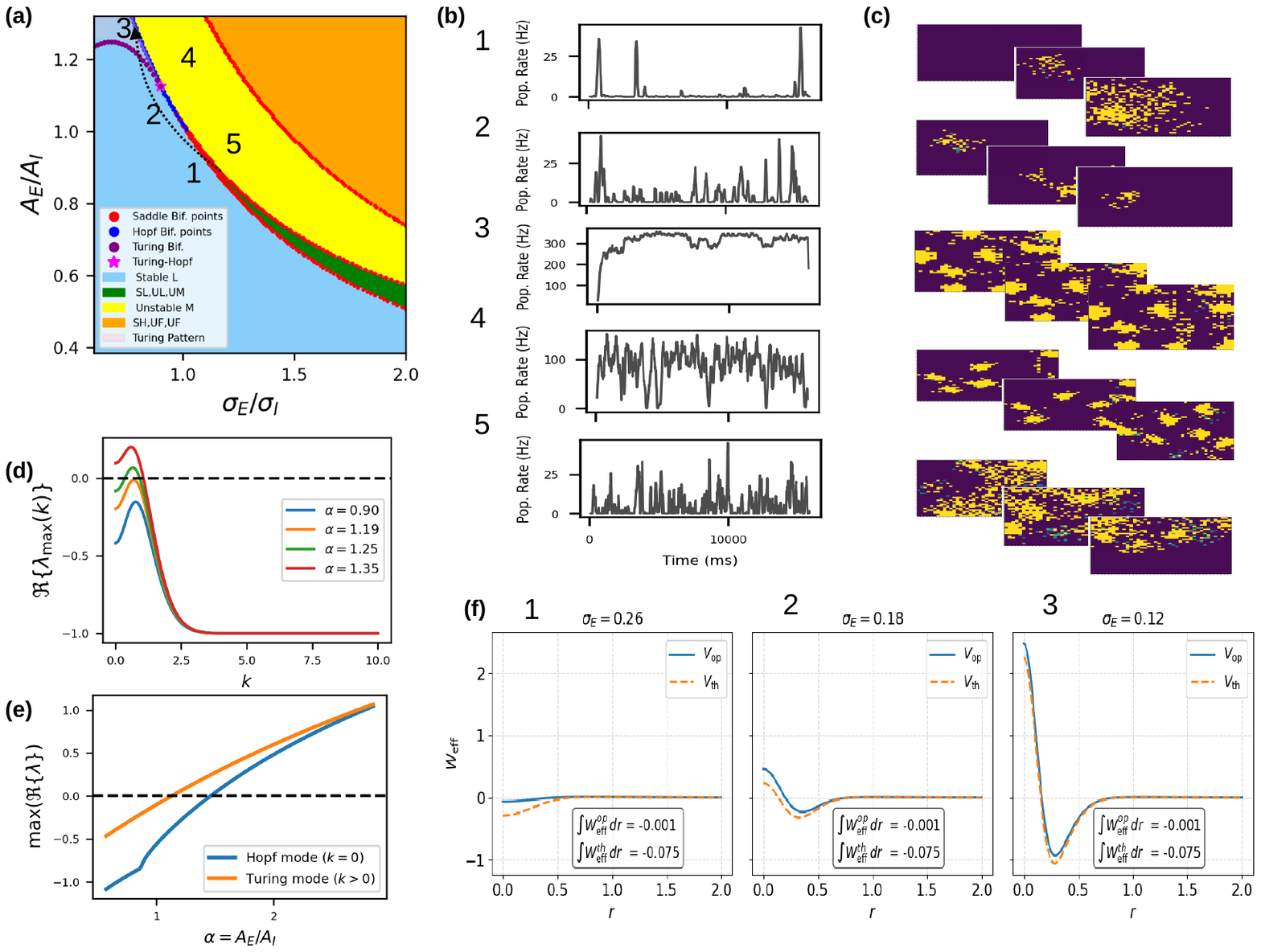}
    \caption{
    \textbf{Turing Bifurcation and Pattern Formation.}
    \textbf{(a)} Bifurcation map illustrating the onset of Turing instabilities alongside saddle-node and Hopf bifurcations of the mean-field equation. The parameter space is defined by the relative kernel widths ($\beta$) and amplitudes ($\alpha$) of the excitatory (Exc.) and inhibitory (Inh.) populations. The Hopf and Turing bifurcation lines intersect at a Hopf-Turing point, where oscillatory patterns emerge. Note that while the exact transition points in the 2D spatial system are shifted relative to the mean-field approximation, the qualitative structure of the map remains valid.
    \textbf{(b)} Network population rates for five representative examples of excitatory weight kernel widths and amplitudes. 
    \textbf{(c)} Snapshots of spatial activity patterns taken at 10~ms intervals. 
    \textbf{(d)} Dispersion curves for a fixed width ratio $\beta = \sigma_E / \sigma_I = 0.75$ and varying amplitude ratios $\alpha = A_E / A_I$. The onset of Turing instability occurs at $\alpha = 1.19$ for the mode $k = 1.24$.
    \textbf{(e)} Thresholds for Turing and Hopf onsets as a function of $\alpha$ at a fixed $\beta$.
    \textbf{(f)} Effective weight kernels for the three networks corresponding to the numbered locations in panels (a) and (b). 
    }
    \label{fig:Turing}
\end{figure}

In spatially extended systems, a homogeneous low-firing state may lose stability through a Turing bifurcation, leading to the emergence of spatial patterns. The relative widths and amplitudes of the excitatory and inhibitory networks are key determinants of pattern formation. Figure~\ref{fig:Turing} shows the Turing bifurcation line in the control-parameter space spanned by the relative kernel widths and amplitudes. In the vicinity of the Turing–Hopf bifurcation ($k \neq 0$) and the Bogdanov–Takens bifurcation point of the mean-field approximation, spontaneous activity takes the form of activity bumps that wander through space with variable size and duration (Fig.~\ref{fig:Turing}b2 and c2). Importantly, the network continues to exhibit scale-free avalanche statistics in this regime, but the avalanches are expressed as spatially diffusive bumps rather than strictly homogeneous bursts. In this case, the relative width of excitatory connections is slightly smaller than inhibition, while the excitatory amplitude is slightly larger. Fig.~\ref{fig:Turing} d and e shows the onset of Turing instability at a fixed width ratio while varying the amplitude ratio. In Appendix~\ref{subsubsec:Bifurcation Analysis}, we further analyze the conditions for a Turing bifurcation ($k \neq 0$), focusing in particular on variations in the relative width of excitation with respect to inhibition while maintaining the balance condition on the integral of the synaptic weights. This corresponds to curved trajectories like the dashed line in Fig.~\ref{fig:Turing}a. Systems numbered 1 to 3 in Fig.~\ref{fig:Turing}a correspond to the effective weight kernels (defined in Appendix~\ref{subsubsec:Effective dynamics})  in Fig.~\ref{fig:Turing}f. Therefore, networks with slightly smaller excitatory kernel widths switch from isotropically propagating avalanches to wandering bumps.

In the following, we initially set the networks in a spatially homogeneous state with similar kernel widths for local excitation and inhibition. However, the connectivity map changes due to synaptic learning during exploration and rest. In Section \ref{sec:online_replay}, we explore the effect of learning a goal-directed path on the 2D connection map and how this change affects the spontaneous activity patterns during rest periods. As shown in Section \ref{sec:diffusive_replay}, the combined effect of STDP and synaptic normalization across the entire manifold during foraging tasks leads to local amplification. This causes the excitatory synaptic kernels to narrow while local weight amplitudes increase, triggering the emergence of spatially structured bursting activity within the network.

\subsection{ Learning an Optimal Path through Online Replay}
\label{sec:online_replay}

The recurrent connectivity is organized such that a localized bump of activity emerges in response to spatially tuned input and moves continuously across the manifold as the agent navigates the environment. In this framework, learning a path toward a goal corresponds to encoding a sequence of place cells together with the actions that transition the agent from a start location to the goal.
When the goal location is invisible or distant from the agent’s starting position, navigation cannot rely on a direct spatial representation of the goal. Instead, the agent must depend on memory of previously experienced trajectories, represented as sequences of locations and associated actions.

We assume a pre-established spatial map, such that place cells are already formed and local state transitions are partially learned through established synaptic weight kernels. However, constructing goal-directed navigation beyond local transitions requires embedding a one-dimensional trajectory within the two-dimensional recurrent manifold. In this section, we examine how the agent refines an initially suboptimal navigation strategy during exploration. Specifically, we consider a network in which the recurrent weights and place-to-action connections encode an indirect path to the goal, and we analyze how forward replay of upcoming place-cell and action representations influences action selection and supports path optimization.

\subsubsection{Formation of a Continuous Line Attractor in 2D  Manifold}

\begin{figure}[hbtp]
    \centering
    \includegraphics[width=1\linewidth]{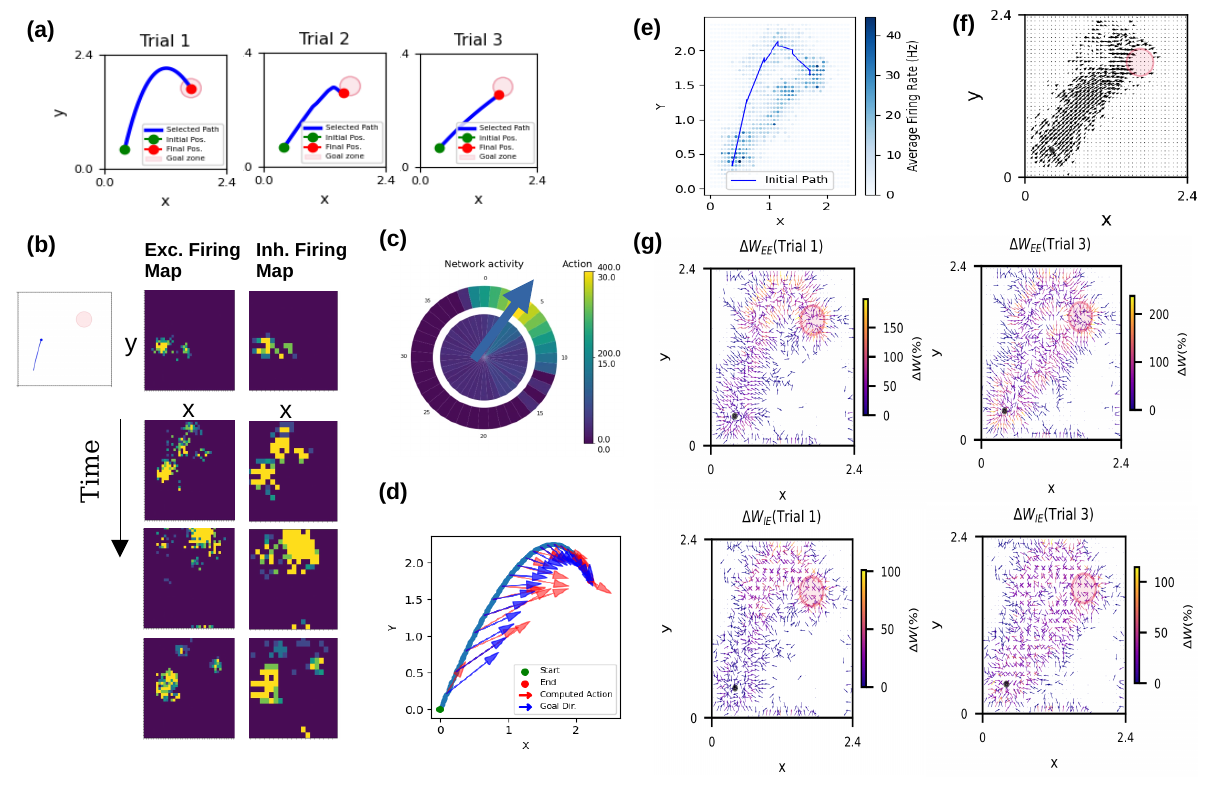}
    \caption{\textbf{Learning a shortcut path via preplay during navigation.}
    \textbf{(a)} Successive trajectories during active exploration: left, the previously learned non-optimal path; middle and right, trajectories selected by the agent using preplay of future states and actions.
    \textbf{(b)} Snapshots of network activity at a location along the non-optimal path (indicated in the left trajectory plot) over a 75~ms time window, showing activity propagating toward the reward location.
    \textbf{(c)} Shift of the activity bump on the ring attractor in the direction corresponding to the vector average of upcoming actions. 
    \textbf{(d)} Action direction at each location along the path, shown while the agent occupies each point for an interval of 75~ms.
    \textbf{(e)} Activity propagation along the newly formed line attractor; the previous attractor is suppressed by the new assembly and downscaled via synaptic normalization.
    \textbf{(f)} Vector average of connections from place cells to action cells, drawn at the spatial location of each cell’s peak firing.
    \textbf{(g)} Maps of recurrent connectivity: (top left) $W_{EE}$ after learning the non-optimal path; (top right) $W_{EE}$ after learning the optimal path; (bottom panels) corresponding maps for $W_{IE}$.}
    \label{fig:online_learning}
\end{figure}

During online exploration, the forward replay of future place cells and action neurons can lead to a smooth integration of the action direction, provided that the width of the connection kernel in the ring attractor is large enough to combine the activity bumps (see Appendix~\ref{subsubsec:Integration_ring_attractor}).
The replay of place–action sequences along a non-optimal path can guide the selection of actions directed toward the goal (Fig.~\ref{fig:online_learning}a-d). As the agent reaches the goal, repeated replay and plasticity induce the formation of a new line attractor within the network. This attractor represents a continuous embedding of the newly learned, goal-directed trajectory within the two-dimensional manifold (Fig.~\ref{fig:online_learning}e–g), leading to gradual and continuous path optimization.

Since the new path is shorter, synaptic plasticity is stronger along it due to reduced decay of the corresponding eligibility traces, leading to modifications of place-to-action connections toward the goal (Fig.~\ref{fig:online_learning}f). Recurrent connections reorganize along the newly formed line attractor through synaptic normalization. Inhibitory-to-excitatory connections ($W_{IE}$) undergo STDP, which automatically adjusts their strength in proportion to excitatory-to-excitatory synapses ($W_{EE}$). As a result, excitatory drive along the newly learned path is naturally balanced by inhibition, suppressing the previous chain of activity and stabilizing the network onto the new attractor (Fig.~\ref{fig:online_learning}g). Consequently, network activity preferentially propagates along the newly optimized, shortest path (Fig.~\ref{fig:online_learning}e).

\subsubsection{Forward and Backward Replay}

A positive plasticity kernel for STDP at short time differences would generate a bidirectional neuronal chain along the learned path. Activity at earlier time and spatial points propagates to assemblies that encode future locations. Due to the dominance of local inhibition, this process results in oscillatory activity within these populations. 

To examine the effects of plasticity along the chain and in neighboring regions, we define an effective weight kernel that captures the linear spatiotemporal relationship between $E(\mathbf{r},t)$ and $I(\mathbf{r},t)$, which governs the dynamics of the slow mode of activity:
\begin{equation}
W_{\mathrm{eff}}(\mathbf{r}) = \hat{W}_{EE}(\mathbf{r}) - \frac{\tilde w_{EI}}{\tilde w_{EE}}\, \alpha\, \hat{W}_{EI}(\mathbf{r}),
\end{equation}
where $\hat{W}_{EE}(\mathbf{r})$ and $\hat{W}_{EI}(\mathbf{r})$ are the normalized weight kernels,  
$\alpha \approx \frac{\tilde w_{IE}}{\tilde w_{II}}$,  
$I(\mathbf{r},t) \approx \alpha\,E(\mathbf{r},t)$,  
and $\tilde w$ is proportional to the maximal amplitude of the corresponding weight kernel (for details, see Appendix~\ref{subsubsec:Effective dynamics}). Figure~\ref{fig:FWD_BWD_Replay}a shows the average radial profile of the effective weight at points along the learned path, both before and after learning through STDP and synaptic normalization, evaluated at the background average membrane potential of excitatory and inhibitory neurons.

When the agent pauses for an extended period at specific locations along a trajectory, both forward and backward replay can occur (Fig.~\ref{fig:FWD_BWD_Replay}b–d). However, longer-lasting inhibitory currents in previously active populations suppress immediate reactivation, biasing the network toward forward replay and the continued progression of activity toward upcoming states.

\begin{figure}[hbtp]
    \centering
    \includegraphics[width=1\linewidth]{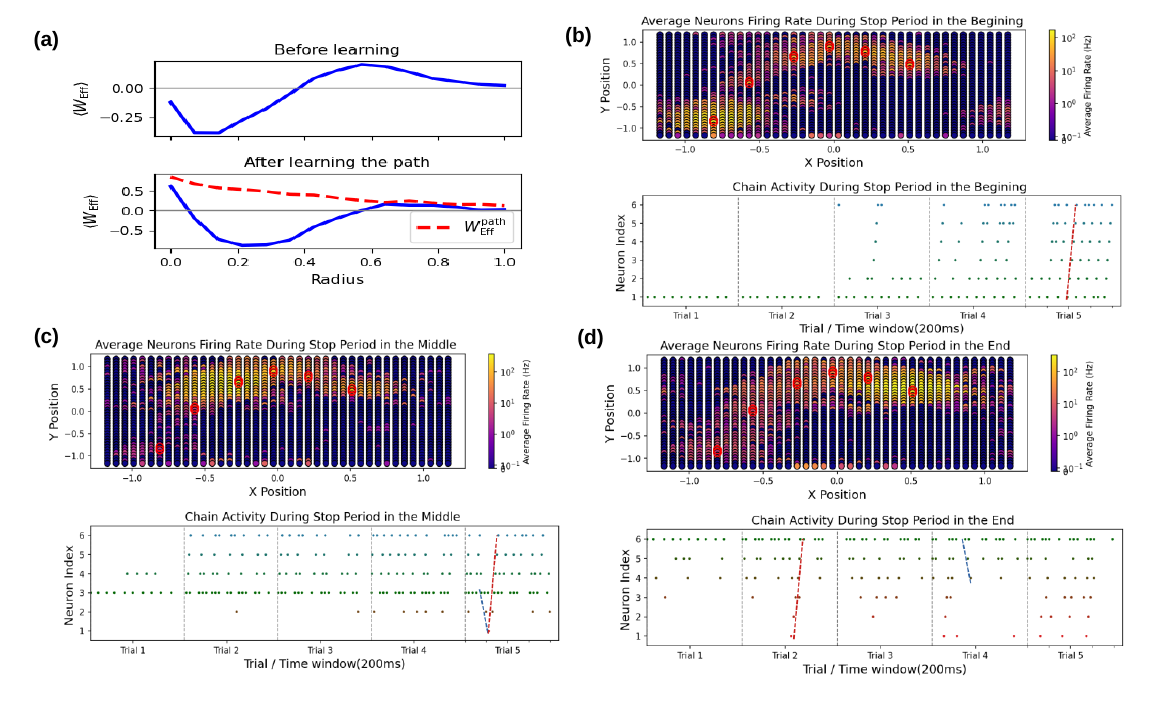}
 \caption{ \textbf{Forward and Backward Replay.}
\textbf{(a)} Effective average synaptic weight from a point along the chain to its neighbors at distance \(r\), \(W_{\mathrm{eff}}(r)\). 
\textbf{(b–d)} Forward and backward wave propagation when the agent is located at the beginning, middle, and end of the path. 
\textit{Top panels:} Average firing rates of neurons when the agent is at a particular location at the end of learning. 
\textit{Bottom panels:} Replay activity of selected neurons along the path during six successive learning trials when the agent is at the corresponding locations. 
Red and blue dashed line indicate samples of forward and backward sequence replay, respectively.
Learning strengthens inner connections and stabilizes replay activity. When the agent is in the middle of the path, forward wave propagation becomes dominant, prioritized, and more frequent.
}
  \label{fig:FWD_BWD_Replay}
\end{figure}

\subsection{ Learning during Offline Replay }
\label{sec:offline_replay}

Here, we discuss the effect of spontaneous activity on the network connections within the 2D manifold, particularly focusing on the potential line attractors that are formed during the early exploration phase. 

\subsubsection{Optimal Potentiation of Neuronal Sequences through Population Bursts}
\begin{figure}[hbtp]
    \centering
    \includegraphics[width=1\linewidth]{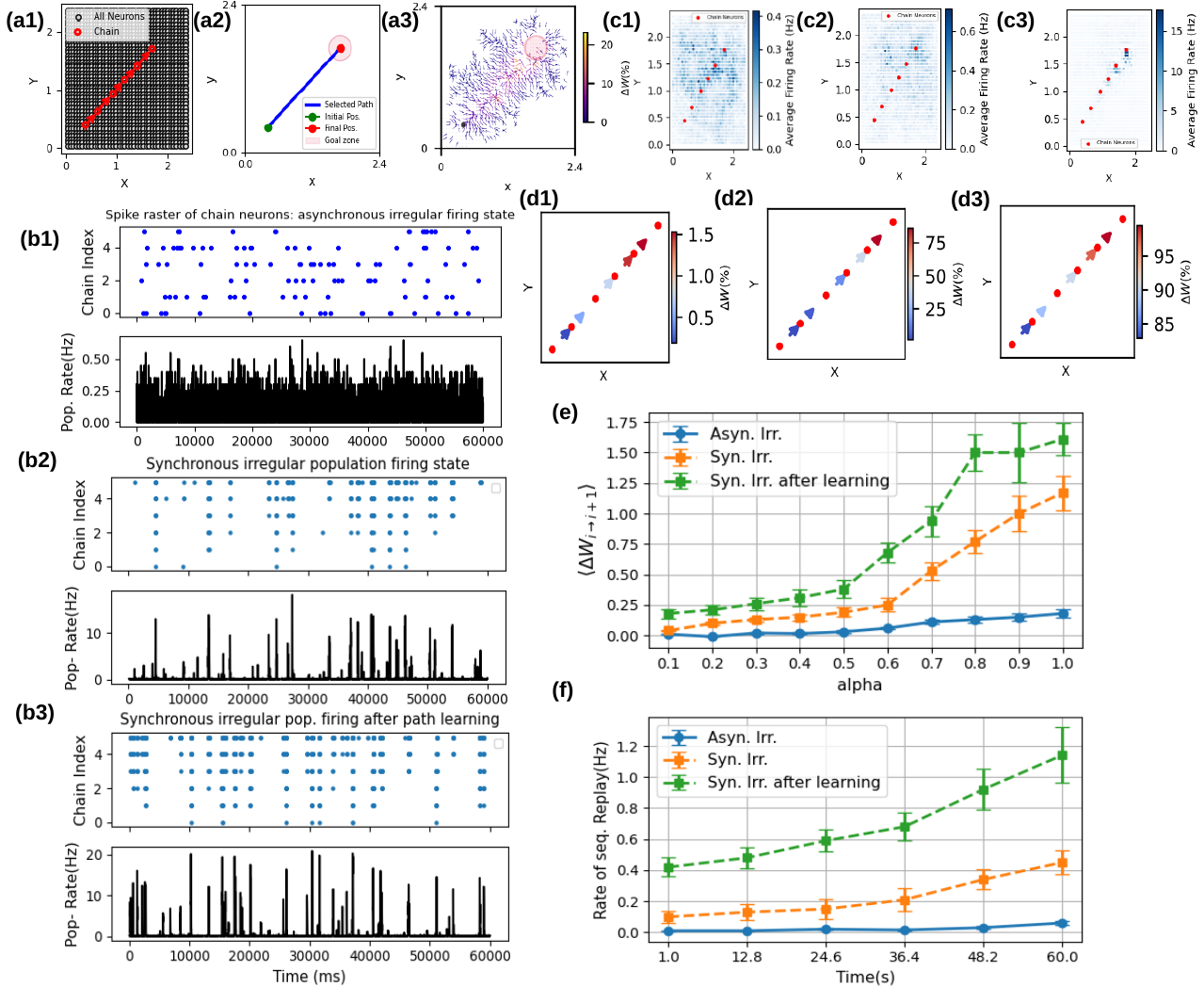}
    \caption{
    \textbf{Influence of Distinct Spontaneous Dynamic Activity States on Sequence Replay} 
    \textbf{(a1)} The sequence embedded as a directed neuronal chain. 
    \textbf{(a2)} Activity propagation during exploration along this path forms an attractor 
    \textbf{(a3)} within the network. 
    \textbf{(b1)} Network activity and spike raster of neurons in the chain when the network is in a homogeneous irregular firing state. 
    \textbf{(b2)} Same for the synchronous irregular (population bursting) state. 
    \textbf{(b3)} Same as (b2), but after the network has undergone synaptic reconfiguration corresponding to (a3) during exploration. 
    \textbf{(c1–3)} Overall heatmaps of neuronal activity for the networks in (b2) and (b3), respectively. 
    \textbf{(d1-3)} Chain weight change (in percentage) for those networks. 
    \textbf{(e)} Change in synaptic weights of a chain with various initial average weights after 30~s of simulation under different network states. 
    \textbf{(f)} Rate of chain activation for a fixed initial normalized weight during 60~s of simulation.
    }
    \label{fig:line_attractor}
\end{figure}

In Fig.~\ref{fig:line_attractor}, we illustrate the effect of spontaneous activity in the form of population bursts on the potentiation of neuronal sequences. As depicted in Fig.~\ref{fig:line_attractor}a1, a directional sequence is constituted by six neurons with medium-range spatial separation of their encoded positions. We aim to analyze the synaptic weight changes along this chain across three distinct scenarios: (i) the chain embedded in a homogeneously connected network characterized by asynchronous firing, (ii) the same chain within a network exhibiting synchronous firing in the form of bursts of varying sizes, and (iii) the chain embedded in a network following exploration, where the local synaptic map has been reorganized to form a line attractor (Fig.~\ref{fig:line_attractor}a2–a3). 

In the asynchronous firing state of a balanced network (Fig.~\ref{fig:line_attractor}b1), inhibitory input effectively decorrelates the activity of pre- and postsynaptic neurons \cite{renart2010asynchronous,decorrelationtetzlaff2012}. This decorrelation minimizes the covariance between their firing rates and thus results in negligible synaptic modification within an embedded chain in the two-dimensional (2D) network(Fig.~\ref{fig:line_attractor}d1).In contrast, during spontaneous activity in the form of population bursts (Fig.~\ref{fig:line_attractor}b2), inhibition shadows excitation with a short temporal delay. This delayed inhibitory feedback generates a cross-covariance function exhibiting a small positive peak at short positive lags,
thereby creating a brief temporal window that favors synaptic strengthening(see Appendix~\ref{subsec:Cross-Correlation}).
Consequently, synapses along the embedded chain experience potentiation during these transient epochs of correlated activity(Fig.~\ref{fig:line_attractor}d2). Furthermore, if the chain is embedded after an exploratory phase of network activity, the synapses connecting adjacent neurons along the path—particularly toward those neurons with the highest firing rates—will be preferentially potentiated.
This selective reinforcement gives rise to a line attractor structure within the 2D network (Fig.~\ref{fig:line_attractor}a3).
As a result, subsequent population bursts more reliably and more frequently activate neurons along the chain (Fig.~\ref{fig:line_attractor}b3),
increasing both the degree of potentiation and its homogeneity within the embedded pathway.

To systematically investigate the effect of the network activity state and chain 
embedding on the potentiation of the chain, we record the weight changes across 
the chain for various values of the initial directional weight. Because the 
networks exhibit different average membrane potential levels—due to variations 
in the combination of external input rates and recurrent connection strengths 
that define their firing regimes—we utilized a normalized $EPSP$ to define the 
initial chain weights $w_0$. Thus, we compare chains with the same initial 
ratio $\alpha = \text{EPSP}(w_0) / \langle \text{EPSP}_{\text{eff}} \rangle$, 
where $\text{EPSP}(w_0)$ is the average $EPSP$ elicited in the postsynaptic 
neuron by a synapse of weight $w_0$, and $\langle \text{EPSP}_{\text{eff}} \rangle$ 
is the average $EPSP$ required to trigger a spike in a postsynaptic neuron 
held at the corresponding mean membrane potential level of different firing regimes.

Fig.~\ref{fig:line_attractor}e shows that the average normalized weight change in the 
synchronous firing regime with an embedded line attractor is higher across all values 
of $\alpha$ after a fixed simulation period, indicating more coordinated firing 
along the chain in this scenario. Higher values of $\alpha$ increase the 
probability of synaptic activation and subsequent modification.
Furthermore, Fig.~\ref{fig:line_attractor}f illustrates the activation rate 
within the embedded pathway across multiple trials for the three scenarios using 
a median initial value of $\alpha = 0.4$, while the synapses undergo weight 
modifications governed by STDP and synaptic normalization keeps the population average synaptic weights in the network constant.

In Appendix \ref{subsec:Cross-Correlation}, we analyze how neuronal input cross-correlation is elevated during population bursts in the ascending phase of avalanches. This elevation creates a window for synaptic modification among the activated cells. This phenomenon contrasts with the inhomogeneous firing regime, where balanced inhibition typically decorrelates neurons (see Fig. \ref{fig:corr_chain} a-b). Furthermore, we demonstrate in Appendix Fig. \ref{fig:corr_chain} c-d how operating within a critical avalanche regime—characterized by a diverging correlation length—enables the network to strengthen connections across its entire extent. Population bursts of varying sizes generate cross-correlations over large spatial distances, facilitating global network plasticity.

\subsubsection{Effect of Offline Replay on Navigation Performance}

\begin{figure}[hbtp]
    \centering
    \includegraphics[width=1\linewidth]{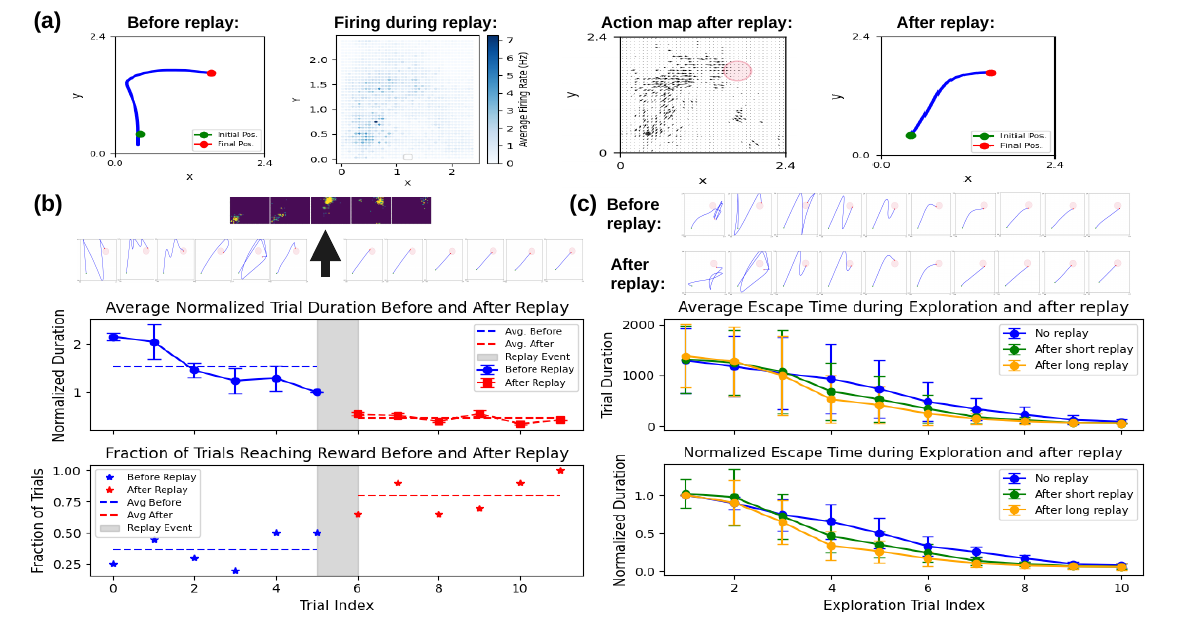}
    \caption{
    \textbf{Effect of replay in reducing latency and stabilizing the path to the goal.}  
    \textbf{(a)} Two exploration trials before and after offline replay. The middle plots show neural activity during rest, indicating increased activity in place cells that encode the previously explored path and demonstrating the effect of replay on the action encoded for each location.
    \textbf{(b)}Average effect of offline replay on latency (top plot) and success rate (bottom plot) during exploration. Results are shown for multiple sequences of explorations, each consisting of six trials starting from the first successful trial in which the agent reaches the goal. During active exploration, internal noise can destabilize the trajectory, causing the agent to miss the goal within the maximum trial duration in subsequent trials. Replay-driven consolidation strengthens the learned path and improves navigation stability.
    \textbf{(c)} Effect of offline replay after different numbers of successful explorations. 
    \textit{(Top)} Example case; \textit{(Middle)} average and variance of trial times across 20 such experiments; 
    \textit{(Bottom)} same as above, but trial times are normalized by the earliest successful exploration trial time.
    }
  \label{fig:offline_replay}
\end{figure}

During periods of rest, external homogeneous noise can trigger the replay of a line attractor, which facilitates a reorganization of the synaptic weights between place cells and action cells. This process enables the local integration of actions across the place cells encoding the learned trajectory. Consequently, after this offline replay period, the network yields a shorter path during the subsequent exploration trial (Fig.~\ref{fig:offline_replay}a).

In Fig.~\ref{fig:offline_replay}b, we illustrate the effect of offline replay on performance following six exploration trials, starting from the first time the agent reaches the reward. As shown both in individual example trials and in the averaged performance across twenty exploration phases, the agent gradually reaches the reward more efficiently over successive trials. Nevertheless, in some fixed-duration trials, the agent occasionally fails to reach the reward, which lowers the overall trial success rate.

During the offline replay period, sequential activity within the recurrent network, as well as the synaptic connections from place cells to action cells, are modified through the combined effects of spike-timing–dependent plasticity (STDP) and synaptic normalization as a form of homeostatic regulation. These mechanisms stabilize the learned sequences of places and actions, improve noise tolerance, and consequently increase the success rate of exploration trials following the rest period.

Specifically, during replay in the rest state, if the response of the ring attractor is sufficiently flexible, the connections between place cells and action cells are adjusted in a coherent manner (see Fig.~\ref{fig:offline_replay}a for a case example of how replay during rest alters the action selection map). Subsequently, during exploration, the firing of place cells simultaneously inputs to all action cells. For each place cell, a bump forms at the smooth average of the upcoming actions, particularly when the action directions vary continuously. However, during exploration, rapid preplay of long action sequences often does not result in a single, stable bump, especially when there is a large separation between action directions. This can hinder optimal action selection over longer trajectories. We further examine the effect of synaptic normalization in the following section to clarify its role in stabilizing network activity during offline replay.

In Fig.~\ref{fig:offline_replay}c, we examine the effect of offline replay after different numbers of successful exploration trials.
When exploration is limited, neither the sequential connections within the recurrent network nor the projections from place cells to action cells are sufficiently strengthened.
Replay events in this case have little effect on the average performance, as they fail to activate behaviorally relevant neuronal sequences.
In contrast, when exploration produces sufficiently strong sequences, replay events can reinforce these connections and even induce some degree of reorganization, thereby reducing both the average latency and its variance.
Finally, when the agent experiences a large number of successful trials during exploration, online learning alone is sufficient to potentiate the sequence and form a stable path, leaving little room for further improvement through offline replay.

\subsubsection{Learning and synaptic normalization Rules Preserve EI Balance and Gain Control to Stabilize Shortcut Solutions}

Plasticity in excitatory-to-inhibitory connections provides the network with a dynamic local balance between excitation and inhibition, while homeostatic synaptic normalization over longer time scales—modeled in our network during offline replay—stabilizes the system globally. Beyond these network-level tuning mechanisms, synaptic plasticity further shapes the attractor landscape by modifying the effective spatial weight kernel around the rewarded path. 

To see the effect of weight change, we define the ratio of synaptic interactions, $\beta(r)$,  as:

\begin{equation}\beta(r) = \dfrac{V_{E(r)E(0)} \cdot V_{I(r)I(0)}}{V_{E(r)I(0)} \cdot V_{I(r)E(0)}}\end{equation}

\begin{figure}[hbtp]
    \centering
    \includegraphics[width=1\linewidth]{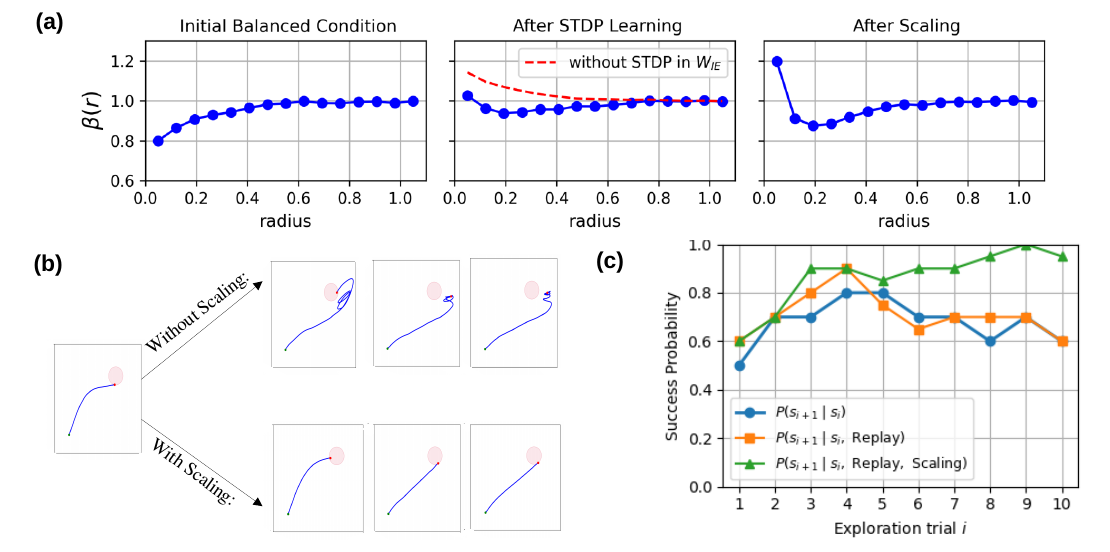}
    \caption{\textbf{Effect of Learning Rules on Network Dynamics and Task Performance.}
    \textbf{(a)} \textbf{(Left)} Radial balance condition before path learning, indicating weak dominance of local inhibition.
    \textbf{(Middle)} Effect of spike-timing--dependent plasticity (STDP) in excitatory-to-inhibitory ($E \to I$) connections on maintaining excitation--inhibition balance.
    \textbf{(Right)} Role of synaptic normalization in shaping a metastable attractor along the rewarded path.
    \textbf{(b)} Exploration behavior following replay periods, with and without synaptic normalization.
    \textbf{(c)} Probability of a successful trial conditioned on the success of the previous trial under three conditions: exploration without excitation--inhibition balance, rest without synaptic normalization, and rest with synaptic normalization.}
    \label{fig:scaling}
\end{figure}

In this expression, $V_{X(r)Y(0)}$ denotes the average postsynaptic potential change, either an EPSP or an IPSP, induced by a neuron of type $Y$ with a place field on the rewarded path onto a neuron of type $X$ with a place field at distance $r$. These values are evaluated at the background membrane potential and averaged over all corresponding neuron pairs along the path.Figure~\ref{fig:scaling}a illustrates this ratio under three distinct conditions: before learning, after synaptic modifications acquired during exploration, and after synaptic normalization during rest. By comparing the products of same-type and cross-type potentials, the ratio $\beta(r)$ provides a measure of how effectively the network maintains a functional balance spatially.

During exploration, in the absence of STDP in excitatory-to-inhibitory (E→I) connections, the network develops a net excitatory region extending over a broad range, which is bounded by inhibitory connections. Subsequent synaptic normalization further refines the synaptic weight kernel by amplifying the weights of more potentiated synapses relative to less potentiated ones, thereby contributing to the formation of a locally quasi-stable bump attractor. Synaptic normalization is also applied to the connections from place cells to action cells, and, together with the tuning of recurrent connections, it helps the network operate in a suitably balanced regime with an embedded quasi-stable line attractor.  The importance of this balance for behavior is shown in Appendix~\ref{subsec:Regimes and Behaviour}: replay in a hyperexcitable regime forms excessive incorrect associations, while in an overinhibited regime it remains local and fails to form long-range associations, so that only the balanced regime substantially improves performance (Fig.~\ref{fig:regime_behaviour}).

Figure~\ref{fig:scaling}b illustrates a sample simulation demonstrating the effect of replay on stabilizing the learned path toward the goal, both without and with synaptic normalization. Figure~\ref{fig:scaling}c shows the probability of successful exploration as a function of the number of previous successful trials under three conditions: (1) without replay and synaptic normalization, (2) with replay but without synaptic normalization, and (3) with both replay and synaptic normalization. When the network operates without synaptic regulation during path embedding, the synaptic weights—despite maintaining overall excitatory–inhibitory balance—continue to grow, producing longer, wandering bursts around the path. This leads to spurious inputs to the action-selection neurons, which can shift the movement direction along the path, induce instability, and consequently reduce performance.

\subsection{Diffusion Dynamics on the 2D Attractor Manifold}
\label{sec:diffusive_replay}

Here, we focus on the pattern of spontaneous activity in a homogeneous 2D attractor model following a random foraging task. The 2D attractor model can be interpreted as an internal cognitive map of the environment, formed through exploration even in the absence of a goal-directed task.
We show that during random foraging, inhomogeneities in synaptic connections increase. In contrast, synaptic rearrangement driven by synaptic normalization and STDP transforms the weight kernel into an effective Mexican-hat profile, causing the network to exhibit spatially structured bursts in the form of wandering activity bumps. 

\subsubsection{Properties of Spontaneous Activity after Foraging Task}
\begin{figure}[hbtp]
    \centering
    \includegraphics[width=1\linewidth]{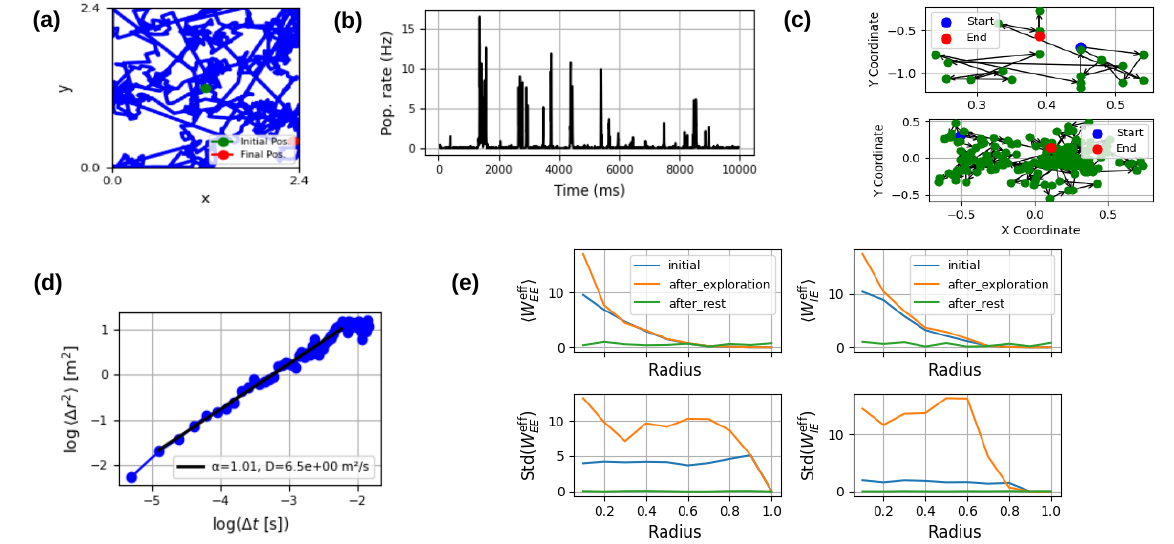}
    \caption{\textbf{Diffusive Offline Replay After Foraging Task}
    \textbf{(a)} Trajectory during the foraging task.
    \textbf{(b)} Population firing organized into bursts of varying sizes.
    \textbf{(c)} Decoded positions during two example bursts. Neural activity is binned in 5\,ms intervals, and positions are decoded using a weighted average of place-cell firing rates.
    \textbf{(d)} Log--log plot of distance versus time for all bursts.
    \textbf{(e)} Mean (top) and standard deviation (bottom) of effective radial weight profiles ($W_{EE}$ on the left and $W_{IE}$ on the right) from a center neuron to its postsynaptic targets at distance $r$, averaged over multiple locations in the environment before foraging, after foraging, and after offline replay with synaptic normalization.
    }
    
    \label{fig:diffusion_char}
\end{figure}

Fig.~\ref{fig:diffusion_char}a shows a random trajectory in an environment with random reward locations which changes each time the agent receives a reward. During exploration STDP would cause synaptic modifications in an inhomogeneous manner based on the random trajectory and random reward locations. we apply further synaptic normalization just at the beginning of the rest trial. Spontaneous activity in absence of spatial input is in the form of population bursts Fig.~\ref{fig:diffusion_char}b but in the form of local bumps which wander around the neuronal network. Binning the spiking data to appropriate bin sizes with regard to bump speed, we decode the animal’s position from population spiking activity. Let $x_i$ denote the center of the place field of neuron $i$, and $n_i(t)$ the number of spikes fired by neuron $i$ within the time bin centered at time $t$. The decoded position $\hat{x}(t)$ was computed using a population vector decoder as
\begin{equation}
\hat{x}(t) = \frac{\sum_i n_i(t)\, x_i}{\sum_i n_i(t)} \, .
\end{equation}
Time bins with no spikes were excluded from the analysis. The decoded position of the bump (Fig.~\ref{fig:diffusion_char}b ) displays characteristics of random Brownian motion, such that the mean squared displacement scales linearly with time ($|\delta x²| \propto \delta t $) (Fig.~\ref{fig:diffusion_char}d). To encode an open field of $2.4 \times 2.4\,\text{m}$, we used place cells with 
spatial fields of width $\sigma = 0.3\,\text{m}$. In accordance with experimental 
observations \cite{wilson1993dynamics}, the network was configured with approximately $50$ neurons 
possessing overlapping place fields at any given location. For this place cell density, the diffusion constants were estimated to lie in the range 
$D \in [3, 10]\,\text{m}^2/\text{s}$. In Appendix~\ref{subsec:Transient Diffusion}, we characterize the fast transient diffusion of activity bumps on a 1D ring network during spontaneous population bursts. This fast transient diffusive dynamics observed in the bursting network may account for the rapid diffusion reported in head direction cell assemblies during REM sleep \cite{Headchaudhuri2019intrinsic}.

We further analyzed the change in the weight profile during spontaneous activity, finding that the random replay strengthens synaptic connections between assemblies representing temporally and spatially proximal locations. This process reduces angular variance in the weight kernel to create a more homogeneous connectivity profile, thereby facilitating the emergence of a coherent, isotropic 2D cognitive map (Fig.~\ref{fig:diffusion_char}e).

In comparison to a homogeneous network with a broad and balanced effective weight kernel 
that exhibits isotropic burst propagation, a network with inhomogeneities and a 
Mexican-hat–shaped profile, arising from synaptic plasticity and synaptic normalization at various locations in the 2D spatial network, shows diffusive bursting dynamics as discussed. In section ~\ref{sec:Balanced_net}, we already discussed the effect of synaptic weight kernel on the population burst. The synaptic normalization of excitatory-to-excitatory ($E \to E$) and excitatory-to-inhibitory ($E \to I$) synaptic weights, while maintaining an overall excitation--inhibition 
balance, transforms the weight kernel from the profile shown in 
Figure~\ref{fig:Turing}f-left (state Nr.1) to the one in 
Figure~\ref{fig:Turing}f-middle (state Nr.2). This structural shift drives the 
network toward a regime of spontaneous activity characterized by wandering local 
bumps as shown in Figure~\ref{fig:Turing}c.

\subsubsection{Effects of Diffusive Replay on Navigation Task}

\begin{figure}[hbtp]
    \centering    \includegraphics[width=\linewidth]{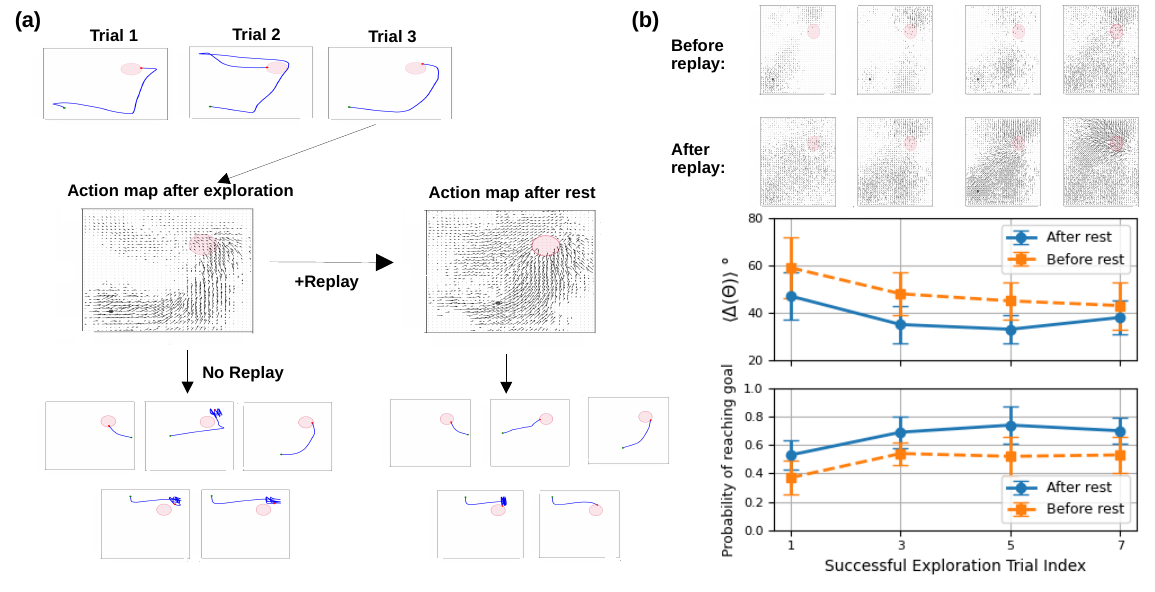}
    \caption{\textbf{Effect of replay in generalization of the place to action map.}\textbf{(a)} The top panel shows three successive exploration trials. The corresponding action maps during these exploration phases are displayed in the second row. In one experiment, the network undergoes diffusive replay, while in the other it does not; the starting point is then changed, and exploration is run for both cases.
    \textbf{(b)} Same experiment as in (a), but performed over ten independent runs. In each run, the effect of replay is compared to no replay after [1, 3, 5, 7] successful exploration trials. }
    \label{fig:Diffusion_Navigation}
\end{figure}

In the case of a goal-directed navigation task, spontaneous activity would not only consist of sequences but, in a network with inherent heterogeneity and spatial structure forming a continuous metastable attractor, it would also appear as diffusive bursts. These bursts diffuse around the learned trajectory, progressively modifying and homogenizing the connections from place cells to action neurons. Figure~\ref{fig:Diffusion_Navigation}a shows that, in the presence of replay in the form of diffusive bursts, the action map generalizes to neighboring locations visited during successful exploration trials. This facilitates more efficient goal-directed navigation from other starting positions in the environment.  

We repeated this experiment across 10 independent runs. In each run, after a number of exploration trials, the network was set to spontaneous firing for a 5 s simulation time, and the success rate of the goal-directed task was measured from variable starting positions, both before and after replay. Figure~\ref{fig:Diffusion_Navigation}b (top) illustrates how diffusive replay homogenizes the action map, as indicated by a reduction in the average angular difference between the encoded action at each place and the actual goal-directed vector. The average and variance of the fraction of successful trials are shown in Fig.~\ref{fig:Diffusion_Navigation}b (bottom).
 Consequently, such diffusive replay facilitates the generalization of the action map and promotes uniformity in local connectivity.

\section{Discussion}

The results presented here establish a dynamical framework for understanding how hippocampal CA3 can facilitate memory consolidation through scale-free population bursting. By situating the network near a Bogdanov–Takens bifurcation, we show that CA3-like dynamics can operate at the edge of both activation and synchronization thresholds, enabling fluid transitions between quiescent states and coordinated activity. Operating in this regime gives rise to power-law burst distributions that satisfy scaling relations consistent with those observed near second-order phase transitions. Unlike traditional models that rely on homogeneous irregular firing, this critical regime maximizes the network’s susceptibility to internal memory traces: because the system is balanced near an instability, even weak synaptic inputs can trigger large-scale recruitment. Crucially, the long spatial correlation lengths that emerge near this transition allow the network to bridge distant neural ensembles, facilitating the activation of sequences that are distributed across the spatial extent of the network rather than confined to local clusters. The stability of this sensitive operating point is maintained through homeostatic regulation via synaptic normalization, ensuring that the network remains responsive to specific synaptic modifications without descending into runaway excitation. Our results further indicate that the propagation of activity is governed by the architecture of the synaptic kernels. In particular, broader local inhibition favors the emergence of wandering activity bumps in the form of a random walk, which prevent global synchronization while allowing activity to sample the learned manifold across multiple spatial and temporal scales.

An important open question concerns the statistical nature of burst dynamics in CA3. While power-law distributed bursts are well documented across cortical areas (\cite{beggs2003neuronal}, \cite{bellay2015irregular}), CA3 bursting activity has been less extensively characterized. Some studies suggest power-law scaling (\cite{timme2016criticality}), whereas others report log-normal size distributions that deviate from strict criticality (\cite{buzsaki2015hippocampal}). This variability highlights the need for further experimental recordings and systematic analyses of the spatiotemporal structure of spontaneous CA3 activity to clarify the mechanisms underlying its burst statistics. Our framework contributes to this discussion by demonstrating how scale-free population bursts can emerge as a direct consequence of excitation–inhibition balance in a network of leaky integrate-and-fire neurons with initially normally distributed synaptic weights. At the same time, the model provides a foundation for further biological refinement. Incorporating log-normal synaptic weight distributions consistent with experimental observations (\cite{buzsaki2014log}) together with intrinsic bursting neuron models could enrich the network’s capacity to generate heavy-tailed dynamics and allow investigation of single-neuron stochastic resonance coordinated by population events. Notably, such skewed weight distributions may themselves arise as emergent properties of ongoing synaptic plasticity interacting with spontaneous activity, suggesting a self-organizing pathway toward the connectivity profiles observed in hippocampal circuits.

A related theoretical question concerns the universality class of the observed critical behavior. Although the critical exponents in our network align with several experimental spike-recording studies \cite{fontenele2019criticality, hahn2010neuronal, jones2023scale}  and share features with oscillatory criticality models \cite{poil2012critical}, the specific exponent values do not coincide with those of standard universality classes typically associated with neural avalanches. Directed percolation, for example, predicts mean-field exponents of $\tau = 1.5$ for burst size and $\alpha= 2.0$ for duration and has been widely proposed as an effective model following the landmark results of Beggs and Plenz (\cite{beggs2003neuronal}). The distinct exponents emerging along the subcritical Hopf bifurcation line in our model suggest that the corresponding critical regime may belong to a different universality class. Clarifying the effective description of this regime will require systematic analyses of finite-size scaling, perturbations, and noise structure, as both internally generated variability and the specific neuronal model can influence the universality class of the dynamics.

Beyond consolidation, our results indicate that sequential memory replay reshapes behaviorally relevant information processing through homogenization and local generalization of the state–action map. In a closed-loop navigation model where the two-dimensional recurrent network serves as a topological embedding of the environment, place-cell sequences and place-to-action mappings acquired during exploration are transformed during both online and offline replay. Replay integrates actions across neurons encoding adjacent spatial locations, enabling the combination of information about past and future trajectories within compressed temporal windows. This mechanism supports path optimization and shortcut discovery by recombining previously experienced trajectories into more efficient routes. Diffusive replay smooths and generalizes the action policy to neighboring locations, effectively interpolating between discrete experiences. In behavioral simulations, this refined action map enables shortcut selection and near-optimal navigation, indicating that diffusive replay constitutes a functional mechanism for transforming episodic traces into a continuous, generalized representation.

The navigation component of the model, however, relies on specific structural assumptions. In particular, the environmental spatial map is pre-embedded in a low-dimensional two-dimensional manifold via state-dependent connectivity, with synaptic strength scaling according to spatial proximity. An important extension would be to examine how such a topological embedding might emerge within a large, sparse recurrent network lacking an initial spatial organization, for example through exploration along one-dimensional trajectories. Integrating grid-cell modules and Dentate Gyrus inputs into an excitation–inhibition balanced CA3 network may prove essential for generating structured spatial representations in this setting. Within a high-dimensional sparse regime, one could then analyze how bursting dynamics influence the efficiency and stability of map formation, and how spontaneous activity primes the network for future navigation.

Moreover, the phenomenological model of action selection assumes that preferred actions are encoded directly in synaptic weights from purely spatial place cells to action-selection neurons, whose activity immediately generates movement commands. While this architecture suffices for goal-directed learning with local state transitions, it becomes restrictive in tasks involving multiple overlapping trajectories, varying starting points, or multiple goal locations. Alternative navigation frameworks using biologically plausible local learning rules that align with temporal difference (TD) learning of a global value function have been shown to implement successor representations \cite{bono2023learning,fang2023neural}, as also formalized in hippocampal predictive-map models \cite{stachenfeld2017hippocampus}. These frameworks decouple reward values from state transitions, allowing more flexible adaptation to changing goals. While the interaction between spontaneous replay and the updating of successor representations has been suggested in both spiking models \cite{bono2023learning} and reinforcement learning models \cite{russek2017predictive}, a fully closed-loop spiking network integrating replay dynamics with an explicit action-selection mechanism has yet to be developed. Such a model could clarify how replay contributes to policy generalization and optimization in more elaborate navigation scenarios.

Finally, the present formulation does not explicitly incorporate the mixed selectivity widely observed in hippocampal neurons \cite{bittner2015conjunctive}. Many cells encode conjunctive combinations of location, direction, context, and goal-related information  \cite{ormond2022hippocampal,howard2014hippocampus,bittner2015conjunctive}, indicating that hippocampal representations are inherently high dimensional \cite{leutgeb2005progressive}. Extending the model to include conjunctive state–action coding or goal-modulated place representations would provide a more biologically grounded substrate for action selection and may enhance the capacity for flexible generalization across tasks and environments.

Together, these considerations outline several directions for future work, aimed at refining the biological realism of the model and deepening our understanding of how critical bursting dynamics and spontaneous replay interact to shape memory consolidation and adaptive navigation.

\section*{Code and Data Availability}
All simulation code was implemented in Python using the NEST simulator and is
available from the corresponding author upon request.

\section*{Acknowledgements}

This work was funded by the Deutsche Forschungsgemeinschaft (DFG, German Research Foundation), grant 397530566 --- FOR 2812, P2.

\bibliographystyle{apalike} 
\bibliography{references}

\newpage
\appendix
\renewcommand{\thefigure}{\thesection.\arabic{figure}}
\setcounter{figure}{0}
\section{Appendix}
\renewcommand{\thesection}{\Alph{section}}
\label{sec:appendix}

\subsection{Network Parameters}
\label{subsec:param_table}
\begin{table}[h!]
\centering
\caption{Model Parameters: Neuronal Dynamics and Modulated Plasticity}
\label{tab:model_parameters}
\scriptsize
\begin{tabularx}{\textwidth}{l l X}
\toprule
Layer / Component & Property & Value / Parameters \\
\midrule
Spatial Input & Neuron Model & \texttt{parrot\_neuron} (Poisson Source) \\
 & Count ($N$) & 441 ($21 \times 21$) \\
 & Connectivity & Rule: Spatial; Kernel: Gaussian ($p_c=1.0, \sigma=0.2$) \\
 & Synapse & \texttt{static\_synapse}; Weight: 3.0 \\
\midrule
Exc. Recurrent Layer & Neuron Model & \texttt{iaf\_cond\_alpha} \\
 & Count ($N$) & 1600 ($40 \times 40$) \\
 & Model Params & $t_{ref}=1.0 \text{ ms}, \tau_{syn}=2.0 \text{ ms}$ \\
 & Exc$\to$Exc / Exc$\to$Inh Conn. & Kernel: Gaussian ($\sigma=0.25$), Sparse: Pairwise-Bernoulli ($p=0.02$) \\
 \cmidrule{2-3}
 & Exc$\to$Exc Synapse & Dopamine-modulated symmetric STDP (Eq.~\ref{eq:sym_kernel}) \\
 & STDP Params & $A_{+} = 0.002$ ($=C_1$), $A_{-} = -A_{+}$, $\tau_{+} = \tau_{-} = 102.0 \text{ ms}$, $W_{max} = 120.0$ \\
 & Modulation & $\tau_c = 200.0 \text{ ms}, \tau_n = 10.0 \text{ ms}$ \\
 & Exc$\to$Exc Weights (Init) & Gaussian; Mean: 10.9, Std: 2.0 \\
 \cmidrule{2-3}
 & Exc$\to$Inh Synapse & Dopamine-modulated symmetric STDP (Eq.~\ref{eq:sym_kernel}) \\
 & STDP Params & $A_{+} = 0.002$ ($=C_1$), $A_{-} = -A_{+}$, $\tau_{+} = \tau_{-} = 102.0 \text{ ms}$, $W_{max} = 140.0$ \\
 & Modulation & $\tau_c = 200.0 \text{ ms}, \tau_n = 10.0 \text{ ms}$ \\
 & Exc$\to$Inh Weights (Init) & Gaussian; Mean: 12.65, Std: 2.0 \\
\midrule
Inh. Recurrent Layer & Neuron Model & \texttt{iaf\_cond\_alpha} \\
 & Count ($N$) & 400 ($20 \times 20$) \\
 & Inh$\to$Exc / Inh$\to$Inh Conn. & Kernel: Gaussian ($\sigma=0.26$) \\
 & Inh$\to$Exc Weights & Gaussian; Mean: -51.5, Std: 5.0 \\
 & Inh$\to$Inh Weights & Gaussian; Mean: -48.0, Std: 5.0 \\
\midrule
Action Layer & Neuron Model & \texttt{iaf\_cond\_alpha} \\
 & Count ($N$) & 40 \\
 & Exc$\to$Act. & Dop.-V-S; $\eta = 0.015, \tau = 60.0 \text{ ms}, \alpha = 1.2, W_{max} = 10.0$ \\
 & Modulation & $\tau_c = 500.0 \text{ ms}, \tau_n = 1.0 \text{ ms}$ \\
\midrule
Border & Neuron Model & \texttt{parrot\_neuron} ($N=8$) \\
 & Synapse & \texttt{static\_synapse}; Weight: -40.0 \\
\bottomrule
\end{tabularx}
\end{table}

\subsection{Action Selection as Ring Attractor Network}
\label{subsec:ring_attractor}

Action neurons represent movement directions on a \textit{ring attractor}, where each neuron \( j \) encodes a preferred direction
\[
\mathbf{a}_j = a_0 (\sin \theta_j, \cos \theta_j),
\]
with
\[
\theta_j = \frac{2\pi j}{N}, \quad N = 40, \quad a_0 = 0.08.
\]

The connectivity follows a periodic \textit{Mexican-hat} profile, characterized by local excitation and broader lateral inhibition. The synaptic weight from neuron \( k \) to neuron \( j \) is defined by the difference of two Gaussian kernels:
\[
w_{jk} = w_E \exp\!\left[-\frac{\Delta\theta_{jk}^2}{2\sigma_E^2}\right] - w_I \exp\!\left[-\frac{\Delta\theta_{jk}^2}{2\sigma_I^2}\right],
\]
where \( w_E > 0 \) and \( \sigma_E \) define the strength and width of local excitation, while \( w_I > 0 \) and \( \sigma_I \) govern the broader inhibition. To ensure the formation of stable localized activity, the inhibitory width is chosen to be broader than the excitatory width (\( \sigma_I > \sigma_E \)). The minimal angular distance between neurons \( j \) and \( k \) is given by
\[
\Delta\theta_{jk} = \mathrm{mod}\!\left(\theta_j - \theta_k + \pi,\, 2\pi\right) - \pi,
\]
which ensures periodic boundary conditions on the ring.

This connectivity structure enables the network to support a stable, localized bump of activity. Neurons with similar preferred directions (\( \Delta\theta_{jk} \approx 0 \)) excite each other strongly, while the broader Gaussian inhibition suppresses distant neurons. This mechanism ensures that only a small subset of tuned neurons remains active, producing smooth and coherent movement trajectories.

The agent's instantaneous action, $\mathbf{a}(t)$, is determined by the population activity of the action neurons. Each neuron $j$ encodes the strength of its preferred action direction, $\mathbf{a}_j$, through its firing rate $\rho_j(t)$. The overall action at time $t$ is computed as the population vector average:
\[
\mathbf{a}(t) = \frac{1}{N} \sum_{j=1}^{N} \rho_j(t) \, \mathbf{a}_j.
\]

\subsubsection{Integration of The Input on the Ring Attractor Network }
\label{subsubsec:Integration_ring_attractor}

The emergence of multiple activity bumps, a single winner-take-all bump, or an integrative bump in the network depends on several key factors: the level of background noise, the strength of external input, the spatial separation of input locations on the ring, and the width of the recurrent weight kernel. In a network with a stable bump solution and strong recurrent connections, activity can be maintained even after the stimulus is removed.

\begin{figure}[hbtp]
    \centering
     \includegraphics[width=\linewidth]{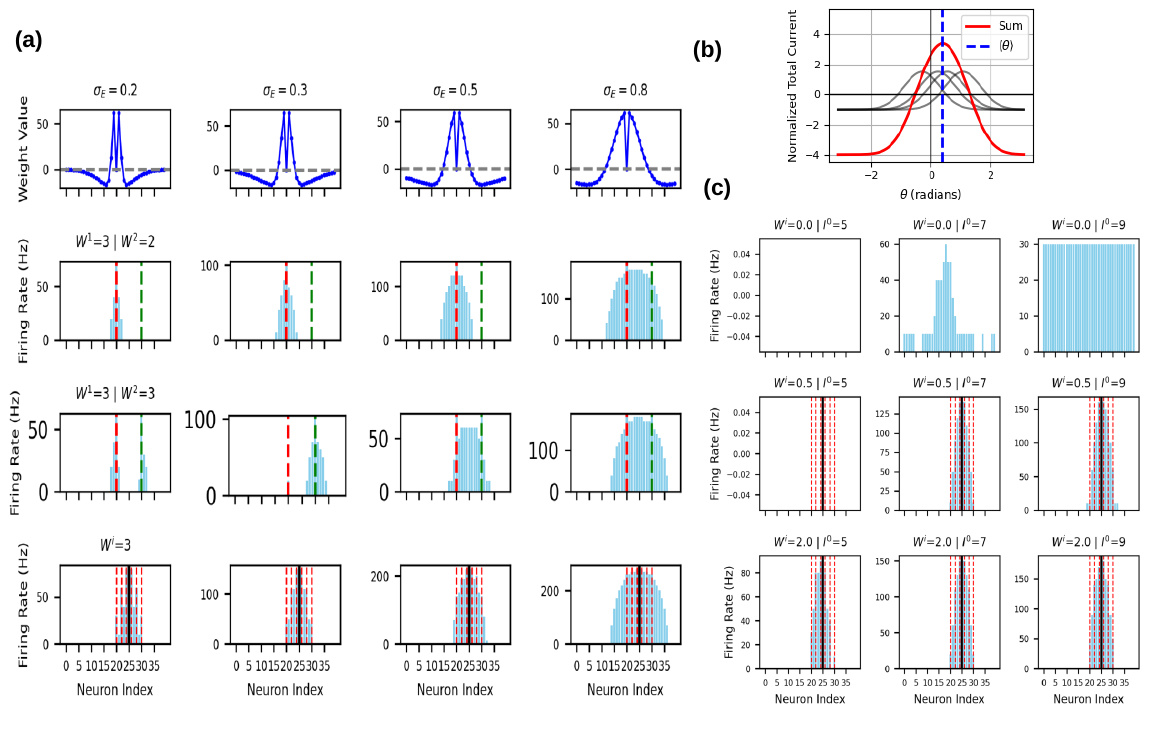}
    \caption{\textbf{Bump formation and integration on the ring attractor network.} (a) Integration, coexistence, and competition of activity bumps on a ring attractor.  
    (b) Top: Network activity in response to two inputs of equal strength applied at a 90° separation. Bottom: Corresponding excitatory connectivity kernels for each network.  
    (c) Illustration of how multiple inputs facilitate bump integration and how input separation affects the width of the integrated bump.  
    (d) Network response to three inputs separated by 90°, presented on top of background noise. A high level of background noise can amplify small input fluctuations and promote the formation of an integrated bump.}
    \label{fig:network-competition}
\end{figure}

We consider a ring network of excitatory populations indexed by phase \( \theta \in [-\pi, \pi) \), with linearized dynamics:

\begin{equation}
\tau_E \frac{\partial E(\theta, t)}{\partial t} = -E(\theta, t) + \alpha \left[ \int_{-\pi}^{\pi} w(\theta - \theta') E(\theta', t) d\theta' + I(\theta, t) \right] + \sigma \eta(\theta, t)
\end{equation}

where the weights follow a periodic Mexican-hat connectivity pattern:

\begin{equation}
w(\theta) = J_E \exp\!\left[-\frac{\Delta\theta^2}{2\sigma_E^2}\right] - J_I \exp\!\left[-\frac{\Delta\theta^2}{2\sigma_I^2}\right], \quad \text{with } \sigma_I > \sigma_E,
\end{equation}

where 
\[
\Delta\theta = \mathrm{mod}(\theta + \pi, 2\pi) - \pi
\]
is the minimal angular distance on the ring.

Consider a ring attractor network where neural activity is defined over an angular variable \( \theta \in [-\pi, \pi] \). Let the network receive external inputs at multiple positions \( \mu_i \), modeled as narrow Gaussians:

\[
I_{\text{ext}}(\theta) = \sum_{i=1}^N A_i \exp\left( -\frac{(\theta - \mu_i)^2}{2 \sigma_{\text{in}}^2} \right), \quad \text{with } \sigma_{\text{in}} \ll \sigma_{\text{rec}}.
\]

The recurrent input is defined as a convolution with a broad, symmetric kernel \( w(\theta) \):

\[
R(\theta) = \int_{-\pi}^{\pi} w(\theta - \theta') u(\theta') \, d\theta'.
\]

At early times when \( u(\theta) \approx 0 \), we approximate the total input as:

\[
I_{\text{eff}}(\theta) = \int w(\theta - \theta') I_{\text{ext}}(\theta') \, d\theta'.
\]

Substituting the expression for \( I_{\text{ext}}(\theta') \):

\[
I_{\text{eff}}(\theta) = \sum_{i=1}^N A_i \int w(\theta - \theta') \exp\left( -\frac{(\theta' - \mu_i)^2}{2 \sigma_{\text{in}}^2} \right) \, d\theta' = \sum_{i=1}^N A_i \int w(\theta - \mu_i - x) \exp\left( -\frac{x^2}{2 \sigma_{\text{in}}^2} \right) \, dx.
\]

Because \( \sigma_{\text{in}} \ll \sigma_{\text{rec}} \), the narrow Gaussian acts like a delta function, and:

\[
I_{\text{eff}}(\theta) \approx \sum_{i=1}^N A_i w(\theta - \mu_i).
\]

To find where the bump forms, we seek:

\[
\theta^* = \arg\max_{\theta} \left( \sum_{i=1}^N A_i w(\theta - \mu_i) \right).
\]

Assume equal amplitudes \( A_i = A \), and \( w \) is a symmetric, unimodal function (e.g., a Gaussian), differentiate with respect to \( \theta \) and set derivative to zero:

\[
\sum_{i=1}^N w'(\theta - \mu_i) = 0.
\]

Since \( w \) is symmetric (i.e., \( w'(-x) = -w'(x) \)), this condition is satisfied at:

\[
\theta^* = \bar{\mu} = \frac{1}{N} \sum_{i=1}^N \mu_i.
\]

Thus, the network bump forms at the mean of the input centers.

\subsubsection*{ Effect of Input Center Variance on Bump Shape}

Let the input centers \( \mu_i \) be sampled from a distribution with variance \( \sigma_{\mu}^2 \).

Then:
\[
I_{\text{eff}}(\theta) \approx \sum_{i=1}^N w(\theta - \mu_i) \approx N \, \mathbb{E}_{\mu}[w(\theta - \mu)].
\]

If \( w \) is a Gaussian of width \( \sigma_{\text{rec}} \), and \( \mu_i \sim \mathcal{N}(\bar{\mu}, \sigma_{\mu}^2) \), then the expected value becomes:

\[
\mathbb{E}_{\mu}[w(\theta - \mu)] = \int w(\theta - \mu) \, \mathcal{N}(\mu; \bar{\mu}, \sigma_{\mu}^2) \, d\mu,
\]

which is the convolution of two Gaussians. Therefore, the effective bump becomes:

\[
I_{\text{eff}}(\theta) \sim \mathcal{N}(\bar{\mu}, \sigma_{\text{eff}}^2), \quad \text{where } \sigma_{\text{eff}}^2 = \sigma_{\text{rec}}^2 + \sigma_{\mu}^2.
\]
Higher variance in input positions (\( \sigma_{\mu}^2 \)) causes the final bump to be broader.
The peak amplitude becomes lower since the bump spreads over a wider domain.
This smoothing is entirely due to the recurrent kernel, not input overlap.

\subsubsection*{Specific Case : Integration of Two Bumps Separated by \( \theta_0 \)}

Assume two narrow inputs centered at \( \mu_1 = -\theta_0/2 \) and \( \mu_2 = +\theta_0/2 \), each of amplitude \( A \). The effective input becomes:

\[
I_{\text{eff}}(\theta) = A \left[ w\left(\theta + \frac{\theta_0}{2}\right) + w\left(\theta - \frac{\theta_0}{2}\right) \right].
\]

To determine whether a single bump forms, examine the curvature of \( I_{\text{eff}}(\theta) \) at the center \( \theta = 0 \):

\[
\left.\frac{d^2 I_{\text{eff}}}{d\theta^2}\right|_{\theta = 0} = A \left[ w''\left( \frac{\theta_0}{2} \right) + w''\left( -\frac{\theta_0}{2} \right) \right] = 2A w''\left( \frac{\theta_0}{2} \right).
\]

If \( w''(\theta_0/2) < 0 \), the center is a maximum and the network integrates the inputs into a single bump. If \( w''(\theta_0/2) > 0 \), the center is a minimum and the inputs remain separate.

For a Gaussian recurrent kernel \( w(\theta) = \exp\left( -\frac{\theta^2}{2\sigma_{\text{rec}}^2} \right) \), we compute:

\[
w''(\theta) = \left( \frac{\theta^2 - \sigma_{\text{rec}}^2}{\sigma_{\text{rec}}^4} \right) w(\theta).
\]

So the sign of the second derivative at \( \theta_0/2 \) depends on whether:

\[
\left( \frac{(\theta_0/2)^2 - \sigma_{\text{rec}}^2}{\sigma_{\text{rec}}^4} \right) < 0 \quad \Rightarrow \quad \theta_0 < 2\sigma_{\text{rec}}.
\]

Therefore, two bumps will merge into one if \( \theta_0 < 2\sigma_{\text{rec}} \); otherwise, the activity remains bimodal.When more than two bumps are present, integration becomes more robust due to overlapping recurrent interactions.

Let \( \mu_1, \dots, \mu_m \sim \mathcal{U}(-\theta_0, \theta_0) \). Then:
\[
I_{\text{eff}}(\theta) \approx A m \cdot \mathbb{E}_{\mu}[w(\theta - \mu)],
\]
which becomes smoother and more Gaussian-like as \( m \) increases, by the Central Limit Theorem.

Thus:
- Even if some bump pairs are separated by \( \theta_{ij} > \theta_c \), the full population can still generate a single broad bump.
- As \( m \to \infty \), the effective input becomes:
\[
I_{\text{eff}}(\theta) \to A m \cdot \left( w * \rho(\theta) \right),
\]
where \( \rho(\theta) \) is the uniform density over the input interval.

 With many input bumps, integration becomes possible even in regimes where pairwise bump distances exceed the critical threshold for merging. The recurrent dynamics average the distributed activity, stabilizing a single coherent bump.The maximum of the effective input always lies at the mean of the input bump locations. With only two or three bumps, integration occurs only when they are sufficiently close. Increasing the number of inputs \( m \) promotes integration by enabling recurrent smoothing of distributed inputs.

\vspace{1em}

\subsection{Dynamics of Balanced Spatial Network Close to Phase Transition}
\label{subsec:Dynamics}
The emergence of power-law scaling in neural activity, often characterized by neuronal avalanches and heavy-tailed burst distributions, suggests that brain might operate near a critical phase transition. To investigate the biophysical conditions under which these power-law bursts arise, we must bridge the gap between microscopic membrane dynamics and macroscopic population stability. In this section, we develop a theoretical framework to analyze the stability of a spatially extended, conductance-based excitatory-inhibitory (EI) network.

We begin by approximating the single-neuron gain function to derive a linearized population model. This allow us to identify the critical points, specifically the Bogdanov-Takens and Turing-Hopf bifurcations, where the system transitions from stable asynchronous firing to structured, scale-free bursting. By examining the stability conditions in Fourier space, we seek to determine how synaptic weights, reversal potentials, and spatial connectivity kernels converge to satisfy the conditions for critical dynamics and robust power-law scaling.

\subsubsection{Single-Neuron Gain Function and Rate Approximation}
\label{subsec:Dynamics_sińgle_neuron_gain}
Consider a single neuron receiving excitatory and inhibitory synaptic inputs as well as external drive. The membrane potential evolves according to a conductance-based integrate-and-fire model:
\begin{equation}
C_m \frac{d V_\alpha}{dt} = - g_{\rm leak} (V_\alpha - V_L) + I_\alpha^{\rm Ext} + I_\alpha^{\rm Rec},
\end{equation}
where $V_\alpha$ is the membrane potential, $g_{\rm leak}$ the leak conductance, and $C_m$ the membrane capacitance. The total external and recurrent synaptic currents are
\begin{align}
I_\alpha^{\rm Ext} &= \sum_\beta G_{\alpha\beta}^{\rm ext} (V_R^\beta - V_\alpha), &
I_\alpha^{\rm Rec} &= \sum_\beta G_{\alpha\beta} (V_R^\beta - V_\alpha),
\end{align}
with $V_R^\beta$ the reversal potential of synapses from population $\beta$, and $G_{\alpha\beta}$ the synaptic conductances. Each conductance evolves according to
\begin{equation}
\tau_{s,\alpha\beta} \frac{d G_{\alpha\beta}}{dt} = -G_{\alpha\beta} + w_{\alpha\beta} r_\beta,
\end{equation}
where $w_{\alpha\beta}$ is the synaptic weight and $r_\beta$ the presynaptic firing rate. 

Neglecting recurrent fluctuations for the mean, the steady-state membrane potential is
\begin{equation}
V_\alpha^* = \frac{
g_{\rm leak} V_L + \sum_\beta w_{\alpha\beta}^{\rm ext} r_{\alpha\beta}^{\rm ext} \tau_{s, \alpha\beta} V_R^\beta
}{
g_{\rm leak} + \sum_\beta w_{\alpha\beta}^{\rm ext} r_{\alpha\beta}^{\rm ext} \tau_{s, \alpha\beta}
}.
\end{equation}
Linearizing around $V_\alpha^*$, the total voltage variance due to stochastic Poisson inputs is
\begin{equation}
\sigma_\alpha^2 = \sum_\beta (V_R^\beta - V_\alpha^*)^2 \frac{w_{\alpha\beta}^2 \, r_\beta \, \tau_{s,\alpha\beta}^2}{2} \frac{\tau_{\rm eff,\alpha}^2}{C_m^2 (\tau_{\rm eff,\alpha} + \tau_{s,\alpha\beta})}, 
\quad
\tau_{\rm eff,\alpha} = \frac{C_m}{g_{\rm leak} + \sum_\beta w_{\alpha\beta} r_\beta \tau_{s,\alpha\beta}}.
\end{equation}

Assuming the membrane potential is approximately Gaussian, $V_\alpha \sim \mathcal{N}(V_\alpha^*, \sigma_\alpha^2)$, the firing rate $r_\alpha$ can be estimated via the rate of threshold crossings (Rice approximation):
\begin{equation}
r_\alpha \approx \frac{1}{2\pi} \frac{\sqrt{\langle \dot V_\alpha^2 \rangle}}{\sigma_\alpha} 
\exp\Big(-\frac{(V_{\rm th} - V_\alpha^*)^2}{2\sigma_\alpha^2}\Big),
\end{equation}
where the variance of the voltage derivative is given by:
\begin{equation}
\langle \dot V_\alpha^2 \rangle = \sum_\beta (V_R^\beta - V_\alpha^*)^2 \frac{w_{\alpha\beta}^2\tau_{s,\alpha\beta}^2 r_\beta}{2 C_m^2} \frac{\tau_{\rm eff}}{(\tau_{\rm eff} + \tau_{s,\alpha\beta})^2}.
\end{equation}

This framework defines an effective \emph{gain function} $F_\alpha(\{r_\beta\})$, mapping input rates to postsynaptic rates. Linearizing this around the operating point allows for the analysis of population firing-rate dynamics:
\begin{equation}
\tau_\alpha \frac{d r_\alpha}{dt} = - r_\alpha + F_\alpha(r_\beta).
\end{equation}

A small change in the conductance $\delta G_{\alpha\beta}$ yields a current perturbation $\delta I_\alpha = \delta G_{\alpha\beta}(V_R^\beta - V_\alpha^*)$. Linearizing the gain function $f_\alpha(I)$ about the operating current $I_\alpha^*$ gives $\delta r_\alpha \approx s_\alpha \delta I_\alpha$, where the sensitivity is:
\begin{equation}
s_\alpha \approx r_\alpha \frac{(V_{\rm th} - V_\alpha^*)}{\sigma_\alpha^2} \frac{1}{g_{\rm tot}^\alpha}.
\end{equation}

\paragraph{Sigmoid Approximation of Gain Functions:}

To capture the input-output relationship in a low-dimensional form, we approximate the population firing rate with a sigmoid function:
\begin{equation}
r_{\rm out} = \frac{R_{\rm max}}{1 + \exp[-(\alpha E + \beta I - \theta)]},
\end{equation}
where $E$ and $I$ are the total excitatory and inhibitory inputs, $\alpha$ and $\beta$ are sensitivities, and $\theta$ is the effective threshold. In a recurrent network, these inputs are proportional to the rates $r_E$ and $r_I$, leading to the self-consistent relaxation dynamics:
\begin{align}
\tau_E \frac{dr_E}{dt} &= -r_E + \frac{R_{\rm max}}{1 + \exp[-(\alpha w_{EE} r_E + \beta w_{EI} r_I - \theta)]}, \\
\tau_I \frac{dr_I}{dt} &= -r_I + \frac{R_{\rm max}}{1 + \exp[-(\alpha' w_{IE} r_E + \beta' w_{II} r_I - \theta')]}.
\end{align}

The parameters $\alpha$, $\beta$, and $\theta$ relate to biophysical scales:
\begin{align}
\alpha &\approx k \, (V_R^{\rm exc} - V_{\rm th}), \\
\beta &\approx k \,  (V_R^{\rm inh} - V_{\rm th}), \\
\theta &\approx - k \left[ g_{\rm syn} \, x_{\rm Ext} \, (V_R^{\rm exc} - V_{\rm th}) + g_L (V_{\rm Leak} - V_{\rm th}) \right].
\end{align}
Here, $k$ is the slope of the logistic fit, typically $0.1$--$1$. This procedure yields a reduced model where weights like $w_{EE}$ serve as control parameters for network bifurcations.

\begin{figure}[hbtp]
    \centering
     \includegraphics[width=\linewidth]{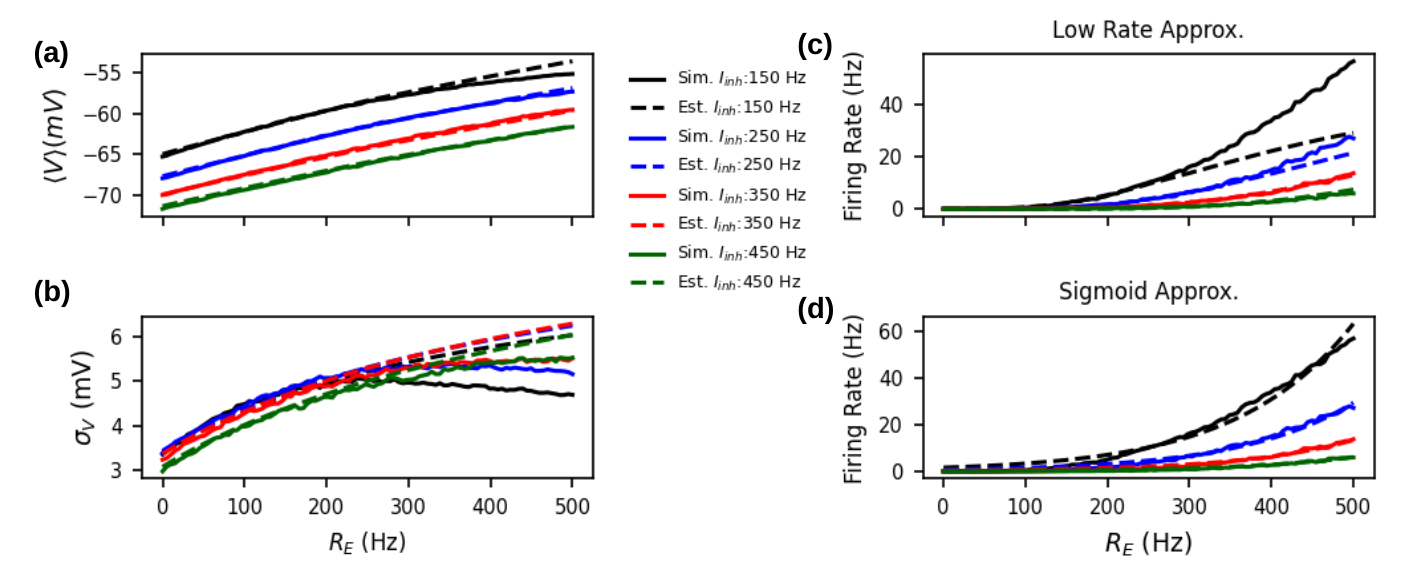}

  \caption{\textbf{Approximation of the neuronal gain function in response to Poisson input.} (a) Average membrane potential, its variance, and firing rate of the neuron model under various fixed external inhibitory rates (each color corresponds to a different inhibition level) and varying excitation. Dashed lines represent analytical results, and solid lines show simulation data.  
(b) Top: Three networks maintained at constant membrane potential by applying balanced excitatory–inhibitory input rates as shown in the top graph (the E–I relationship is linear).  
Middle: Membrane potential variance under balanced input; at higher balanced rates, both variance and firing rate decrease.  
Bottom: Output rate and self-consistency curve for balanced homogeneous solutions in the network. }
\end{figure}

\subsubsection{Bifurcation Analysis of Spatial EI network}
\label{subsubsec:Bifurcation Analysis}

Consider a spatially extended recurrent network of excitatory ($E$) and inhibitory ($I$) neurons. The population firing rate at location $x$ for population $\alpha \in \{E,I\}$ is determined by a general gain function:
\begin{equation}
r_\alpha(x,t) = F_\alpha(E_\alpha(x,t), I_\alpha(x,t)),
\end{equation}
where the total excitatory ($E_\alpha$) and inhibitory ($I_\alpha$) inputs are defined as:
\begin{align}
E_\alpha(x,t) &= \int w_{\alpha E} W_{\alpha E}(x-x') r_E(x',t)\,dx' + x_\alpha^{\rm ext, exc}, \\
I_\alpha(x,t) &= \int w_{\alpha I} W_{\alpha I}(x-x') r_I(x',t)\,dx' + x_\alpha^{\rm ext, inh}.
\end{align}

Introducing first-order relaxation dynamics with population time constants $\tau_\alpha$ gives:
\begin{equation}
\tau_\alpha \frac{\partial r_\alpha(x,t)}{\partial t} = - r_\alpha(x,t) + F_\alpha(E_\alpha(x,t), I_\alpha(x,t)).
\end{equation}

\medskip
\noindent
\textbf{Linearization around the low-firing operating point:}  
Let $r_\alpha^*$ denote the spatially uniform steady-state rate, with the corresponding operating point defined by the mean membrane potential $V_\alpha^*$ and voltage variance $\sigma_\alpha^2$. We define small perturbations:
\begin{equation}
\delta r_\alpha(x,t) = r_\alpha(x,t) - r_\alpha^*.
\end{equation}

Linearizing the gain function around the operating point yields:
\begin{equation}
\delta r_\alpha(x,t) \approx s_\alpha \, \delta I_\alpha(x,t),
\qquad
s_\alpha := \left. \frac{\partial F_\alpha}{\partial I_\alpha} \right|_{I_\alpha^*},
\end{equation}
where $\delta I_\alpha(x,t)$ is the perturbation in the total input current. For conductance-based synapses, this perturbation is:
\begin{equation}
\delta I_\alpha(x,t) = \sum_\beta (V_R^\beta - V_\alpha^*) \, \delta G_{\alpha\beta}(x,t),
\quad
\delta G_{\alpha\beta}(x,t) = w_{\alpha\beta} \int W_{\alpha\beta}(x-x') \delta r_\beta(x',t) \, dx'.
\end{equation}

By defining the effective operating-point weights as $\widetilde w_{\alpha\beta} := w_{\alpha\beta} (V_R^\beta - V_\alpha^*)$, the linearized dynamics for the spatial network become:
\begin{equation}
\tau_\alpha \frac{\partial \delta r_\alpha(x,t)}{\partial t} = - \delta r_\alpha(x,t) + \sum_\beta s_\alpha \int \widetilde w_{\alpha\beta} W_{\alpha\beta}(x-x') \delta r_\beta(x',t) \, dx'.
\end{equation}

\medskip
\noindent
\textbf{Fourier-space representation:}  
Applying a spatial Fourier transform $\delta r_\alpha(k,t) = \int \delta r_\alpha(x,t) e^{-i k x} dx$ yields:
\begin{equation}
\tau_\alpha \frac{\partial \delta r_\alpha(k,t)}{\partial t} = - \delta r_\alpha(k,t) + \sum_\beta s_\alpha \widetilde w_{\alpha\beta} \widehat W_{\alpha\beta}(k) \delta r_\beta(k,t),
\end{equation}
where $\widehat W_{\alpha\beta}(k)$ is the Fourier transform of the spatial connectivity kernel. The Jacobian matrix $J(k)$ for mode $k$ has components:
\begin{align}
J_{\alpha\beta}(k) &= \frac{s_\alpha}{\tau_\alpha} \widetilde w_{\alpha\beta} \widehat W_{\alpha\beta}(k) \quad (\alpha \neq \beta), \\
J_{\alpha\alpha}(k) &= -\frac{1}{\tau_\alpha} + \frac{s_\alpha}{\tau_\alpha} \widetilde w_{\alpha\alpha} \widehat W_{\alpha\alpha}(k).
\end{align}

The stability of mode $k$ is determined by the characteristic equation $\lambda^2 - \operatorname{Tr} J(k) \lambda + \det J(k) = 0$, where:
\begin{align}
\operatorname{Tr}J(k) &= -\frac{1}{\tau_E} - \frac{1}{\tau_I} + \frac{s_E}{\tau_E}\widetilde w_{EE}\widehat{W}_{EE}(k) + \frac{s_I}{\tau_I}\widetilde w_{II}\widehat{W}_{II}(k), \\
\det J(k) &= \left(-\frac{1}{\tau_E} + \frac{s_E}{\tau_E}\widetilde w_{EE}\widehat{W}_{EE}(k)\right) \left(-\frac{1}{\tau_I} + \frac{s_I}{\tau_I}\widetilde w_{II}\widehat{W}_{II}(k)\right) \nonumber \\
&\quad - \frac{s_E s_I}{\tau_E \tau_I}\widetilde w_{EI}\widetilde w_{IE}\widehat{W}_{EI}(k)\widehat{W}_{IE}(k).
\end{align}
Turing-like instabilities (spatial patterning) occur when $\det J(k) < 0$ for some $k > 0$, while Hopf bifurcations (oscillations) occur when $\operatorname{Tr} J(k) > 0$ with $\det J(k) > 0$.

\paragraph{Homogeneous Mode ($k=0$): Global Bifurcations:}

For the homogeneous (spatially uniform) perturbation $k=0$, we have $\widehat{W}_{\alpha\beta}(0)=1$. The Jacobian reduces to:
\begin{equation}
J(0)=
\begin{pmatrix}
-\dfrac{1}{\tau_E} + \dfrac{s_E}{\tau_E}\,\widetilde{w}_{EE} & \dfrac{s_E}{\tau_E}\,\widetilde{w}_{EI} \\[8pt]
\dfrac{s_I}{\tau_I}\,\widetilde{w}_{IE} & -\dfrac{1}{\tau_I} + \dfrac{s_I}{\tau_I}\,\widetilde{w}_{II}
\end{pmatrix},
\end{equation}
and the trace and determinant become:
\begin{align}
\operatorname{Tr}J(0) &= -\frac{1}{\tau_E} - \frac{1}{\tau_I} + \frac{s_E}{\tau_E}\widetilde{w}_{EE} + \frac{s_I}{\tau_I}\widetilde{w}_{II}, \\
\det J(0) &= \left(-\frac{1}{\tau_E} + \frac{s_E}{\tau_E}\widetilde{w}_{EE}\right) \left(-\frac{1}{\tau_I} + \frac{s_I}{\tau_I}\widetilde{w}_{II}\right) - \frac{s_E s_I}{\tau_E \tau_I}\,\widetilde{w}_{EI}\widetilde{w}_{IE}.
\end{align}

A saddle-node bifurcation occurs when $\det J(0) = 0$. Solving explicitly for the required inhibitory-to-excitatory effective weight $\widetilde{w}_{EI}$ (or $\widetilde{w}_{IE}$) yields:
\begin{equation}
\widetilde{w}_{IE} = \frac{(s_E \widetilde{w}_{EE} - 1)(s_I \widetilde{w}_{II} - 1)}{s_E s_I \widetilde{w}_{EI}}.
\end{equation}

A Hopf bifurcation occurs when the eigenvalues cross the imaginary axis ($\operatorname{Tr} J(0) = 0$ and $\det J(0) > 0$). This condition is satisfied when:
\begin{equation}
\frac{s_E}{\tau_E}\widetilde{w}_{EE} + \frac{s_I}{\tau_I}\widetilde{w}_{II} = \frac{1}{\tau_E} + \frac{1}{\tau_I}.
\end{equation}

Since $s_\alpha$, $\tau_\alpha$, and $\widetilde{w}_{\alpha\beta}$ all depend implicitly on the fixed point $V_\alpha^*$, the saddle-node and Hopf loci in the $(w_{EE}, w_{IE})$ plane are \emph{implicit curves} that must be solved self-consistently.

A Bogdanov-Takens (BT) point at $k=0$ requires the simultaneous vanishing of the trace and determinant:
\begin{align}
\label{BT-det-tilde}
\left(-\frac{1}{\tau_E} + \frac{s_E}{\tau_E}\widetilde{w}_{EE}\right) \left(-\frac{1}{\tau_I} + \frac{s_I}{\tau_I}\widetilde{w}_{II}\right) &= \frac{s_E s_I}{\tau_E \tau_I}\,\widetilde{w}_{EI}\widetilde{w}_{IE}, \\
\label{BT-tr-tilde}
\frac{s_E}{\tau_E}\widetilde{w}_{EE} + \frac{s_I}{\tau_I}\widetilde{w}_{II} &= \frac{1}{\tau_E} + \frac{1}{\tau_I}.
\end{align}

Because $\widetilde{w}_{\alpha\beta}=w_{\alpha\beta}(V_R^\beta - V_\alpha^*)$, these relations depend on the operating voltages $V_\alpha^*$, the fluctuation amplitudes $\sigma_\alpha$, and the mean recurrent conductances. Parameter variations that shift the operating point (e.g., changes in external drive $r_{\rm ext}$) modify all coefficients in the BT equations.

In the limit of nearly identical populations ($s_E \approx s_I = s$ and $\tau_E \approx \tau_I = \tau$), the BT relations simplify. From the trace equation, we obtain $s(\widetilde{w}_{EE} + \widetilde{w}_{II}) = 2$. If the gain $s$ is sufficiently large such that the terms of order $s^2$ dominate, the determinant relation reduces to a simple product-balance:
\begin{equation}
\widetilde{w}_{EE}\widetilde{w}_{II} \approx \widetilde{w}_{EI}\widetilde{w}_{IE}.
\end{equation}

If $s_\alpha$ is not large or if there is significant asymmetry between populations, the full expression for the determinant must be used. Rewriting \eqref{BT-det-tilde} reveals the deviation from the product-balance:
\begin{equation}
\widetilde{w}_{EE}\widetilde{w}_{II} - \widetilde{w}_{EI}\widetilde{w}_{IE} = \frac{1}{s_E s_I}\left(\frac{s_I \widetilde{w}_{II}}{\tau_E} + \frac{s_E \widetilde{w}_{EE}}{\tau_I} - \frac{1}{\tau_E \tau_I}\right).
\end{equation}

\paragraph{Numerical Bifurcation Analysis:}

To facilitate numerical exploration of the network's phase space, we apply the sigmoid gain function $F_\alpha(I) = R_{\max} / (1 + e^{-(I - \theta_\alpha)})$. The sensitivity (gain slope) $s_\alpha$ is state-dependent and is determined by the output firing rate:
\begin{equation}
s_\alpha = \frac{dF_\alpha}{dI} = \frac{1}{R_{\max}} r_\alpha^* (R_{\max} - r_\alpha^*).
\end{equation}

In the \textbf{low-rate approximation} ($r_\alpha^* \ll R_{\max}$), which corresponds to the fluctuation-driven regime observed in vivo, the gain simplifies to a linear dependency:
\begin{equation}
s_\alpha \approx r_\alpha^*.
\end{equation}

The bifurcation boundaries in the $(w_{EE}, w_{IE})$ plane are determined by the roots of the Jacobian characteristic equation at $k=0$. Because $s_\alpha$ depends on the fixed point $r_\alpha^*$, the following numerical procedure is employed:
For a given set of synaptic weights, the self-consistent rates $r_E^*$ and $r_I^*$ are found by solving the transcendental equations:
\begin{align}
r_E^* &= \frac{R_{\max}}{1 + \exp[-(\alpha w_{EE} r_E^* + \beta w_{EI} r_I^* - \theta_E)]} \\
r_I^* &= \frac{R_{\max}}{1 + \exp[-(\alpha' w_{IE} r_E^* + \beta' w_{II} r_I^* - \theta_I)]}
\end{align}

The locus of saddle-node bifurcations is found where $\det J(0) = 0$. Substituting the sigmoid sensitivities, this boundary satisfies:
\begin{equation}
\left(1 - \alpha w_{EE} s_E\right)\left(1 - \beta' w_{II} s_I\right) - \beta w_{EI} s_E \alpha' w_{IE} s_I = 0.
\end{equation}

The Hopf bifurcation curve is localized where $\operatorname{Tr} J(0) = 0$, provided $\det J(0) > 0$. This requires:
\begin{equation}
\frac{\alpha w_{EE} s_E - 1}{\tau_E} + \frac{\beta' w_{II} s_I - 1}{\tau_I} = 0.
\end{equation}

The intersection of these two curves defines the Bogdanov-Takens (BT) point. Near this point, the system is highly susceptible to noise, and the interplay between the vanishing trace and determinant leads to the emergence of the scale-free power-law bursts described in Section 3.2. This numerical approach allows us to map the "avalanche regime" specifically to the biophysical parameters of the CA3 recurrent network. For fixed values of parameters as listed in Table \ref{tab:model_parameters} , the low Bogdanov-Takens (BT) 
point with the Sigmoid approximation lies at:
\[
W_{IE}^{\mathrm{eff}} = W_{IE} \left(V_R^E - V_{\mathrm{th}}\right) = 0.11 \quad \text{and} \quad 
W_{EE}^{\mathrm{eff}} = W_{EE} \left(V_R^E - V_{\mathrm{th}}\right) = 0.13.
\]
The values calculated using the low-firing fixed point are $W_{IE}^{\mathrm{eff}} = 0.10$ 
and $W_{EE}^{\mathrm{eff}} = 0.12$. Based on an extensive grid search of simulations 
in locally connected and random networks, the values for the BT point in the 
$[W_{EE}^{\mathrm{eff}}, W_{IE}^{\mathrm{eff}}]$ plane are $[0.09, 0.09]$ and $[0.10, 0.115]$, respectively. The BT point differs across methods because of approximation, network structure, and noise. The sigmoid approximation is smoother and works well for mapping the entire bifurcation diagram, but the calculation using the low-firing rate approximation is more accurate for the actual BT point location. In simulations, local connections amplify recurrent effects, so smaller effective weights are sufficient to reach the BT point, whereas random networks require slightly higher weights.

\paragraph{Beyond the Homogeneous Mode: Turing Instability under Varying Excitatory Width}

We consider a spatially extended excitatory-inhibitory (EI) network with Gaussian connectivity kernels in 2D. To isolate the effects of spatial scales, we assume:
\begin{equation}
\sigma_{EE} = \sigma_{IE} = \sigma_E, \qquad \sigma_{EI} = \sigma_{II} = \sigma_I,
\end{equation}
with the width ratio $\sigma_E = \beta \sigma_I$. To keep the total synaptic drive constant as $\beta$ varies, we fix the spatial integral of the effective weights $\widetilde{w}_{\alpha\beta}(x) = w_{\alpha\beta}(x)(V_R^\beta - V_\alpha^*)$:
\begin{equation}
\mathcal{I}_{\alpha\beta} = \int_{\mathbb{R}^2} \widetilde{w}_{\alpha\beta}(x) \, d^2x = \text{const.}
\end{equation}
For a 2D Gaussian kernel $\widetilde{w}_{\alpha\beta}(r) = \widetilde{W}_{\alpha\beta} \exp(-r^2/2\sigma_{\alpha\beta}^2)$, this implies $\widetilde{W}_{\alpha\beta} = \mathcal{I}_{\alpha\beta} / (2 \pi \sigma_{\alpha\beta}^2)$. Consequently, narrowing the spatial reach of excitation (decreasing $\beta$) increases its peak weight to preserve the $k=0$ mode stability.

The system is considered close to a global bifurcation where the homogeneous state is marginally stable ($\lambda_{\max}(k=0) \approx 0$). The Fourier transform of the kernels is $\widehat{W}_{\alpha\beta}(k) = \exp(-|k|^2 \sigma_{\alpha\beta}^2 / 2)$. The mode-dependent Jacobian $J(k)$ is:
\begin{equation}
J(k) =
\begin{pmatrix}
-\frac{1}{\tau_E} + \frac{s_E}{\tau_E} \widetilde{W}_{EE} \widehat{W}_{E}(k) & \frac{s_E}{\tau_E} \widetilde{W}_{EI} \widehat{W}_{I}(k) \\
\frac{s_I}{\tau_I} \widetilde{W}_{IE} \widehat{W}_{E}(k) & -\frac{1}{\tau_I} + \frac{s_I}{\tau_I} \widetilde{W}_{II} \widehat{W}_{I}(k)
\end{pmatrix}.
\end{equation}

A Turing instability occurs if $\det J(k) < 0$ for some $k > 0$. Expanding $\det J(k)$ for small $q = |k|^2$, the condition for the emergence of a spatial instability is $\frac{d}{dq} \det J(q) |_{q=0} < 0$. Using the fixed integral condition $\widetilde{W}_{EE} = \mathcal{I}_{EE}/(2\pi \beta^2 \sigma_I^2)$ and $\widetilde{W}_{II} = \mathcal{I}_{II}/(2\pi \sigma_I^2)$, the inequality becomes:
\begin{equation}
\frac{\Delta \mathcal{I}}{\sigma_I^2} \frac{\beta^2 + 1}{\beta^2} > 2 \pi \left( \frac{\tau_I}{s_I} \mathcal{I}_{EE} + \frac{\tau_E}{s_E} \mathcal{I}_{II} \right),
\end{equation}
where $\Delta \mathcal{I} = \mathcal{I}_{EE} \mathcal{I}_{II} - \mathcal{I}_{EI} \mathcal{I}_{IE}$. 

As $\beta$ decreases, the factor $(\beta^2 + 1)/\beta^2$ increases. This demonstrates that making excitation more spatially localized (while maintaining the same total synaptic weight) destabilizes the homogeneous mode in favor of finite-$k$ spatial patterns. This corresponds to curved trajectories like the dashed line in Fig.~\ref{fig:Turing}a in the main text.

\subsubsection{Effective Dynamics near Phase Transition: Slow--Fast Decomposition and Transient Bursting}
\label{subsubsec:Effective dynamics}

The network operates in a parameter regime where both a saddle-node and a Hopf bifurcation are nearby in parameter space. We focus on the vicinity of a subcritical Hopf bifurcation associated with the high-activity fixed point, which renders this state unstable. In this regime, a saddle point separates the quiescent fixed point from the unstable high-activity branch created at the saddle-node bifurcation. Although the Hopf bifurcation determines the stability of the high-activity state, the transient dynamics relevant for avalanche-like events are governed by stochastic trajectories in the neighborhood of the saddle. The presence of the saddle-node introduces a single near-marginal direction, leading to a natural separation of timescales. We therefore employ a slow–fast decomposition controlled by proximity to the saddle-node bifurcation, which captures the transient behavior underlying avalanches.

We consider small perturbations of excitatory and inhibitory population rates, $\delta r_E(\mathbf{k},t)$ and $\delta r_I(\mathbf{k},t)$, around a homogeneous steady state. Linearizing the rate equations and transforming to Fourier space yields, for each wavenumber $k = |\mathbf{k}|$:
\begin{equation}
\tau_\alpha \frac{\partial}{\partial t}\,\delta r_\alpha(k,t) = -\delta r_\alpha(k,t) + s_\alpha \sum_\beta \widetilde w_{\alpha\beta}\,\widehat{W}_{\alpha\beta}(k)\,\delta r_\beta(k,t), \qquad \alpha,\beta \in \{E,I\}.
\end{equation}

The Jacobian $J(k)$ is given by:
\begin{equation}
J(k)=
\begin{pmatrix}
-\dfrac{1}{\tau_E} + \dfrac{s_E}{\tau_E}\,\widetilde w_{EE}\,\widehat{W}_{EE}(k) &
\dfrac{s_E}{\tau_E}\,\widetilde w_{EI}\,\widehat{W}_{EI}(k)\\[8pt]
\dfrac{s_I}{\tau_I}\,\widetilde w_{IE}\,\widehat{W}_{IE}(k) &
-\dfrac{1}{\tau_I} + \dfrac{s_I}{\tau_I}\,\widetilde w_{II}\,\widehat{W}_{II}(k)
\end{pmatrix}.
\end{equation}

Let $\lambda_{1,2}(k)$ denote the eigenvalues of $J(k)$. Close to a \emph{saddle-node bifurcation}, where one eigenvalue of $J(0)$ approaches zero, the spectrum separates into a slow and a fast mode:
\begin{equation}
\Re[\lambda_{\text{slow}}(k)] \approx 0^{-}, \qquad \Re[\lambda_{\text{fast}}(k)] \ll 0.
\end{equation}

In this regime, the fast subsystem relaxes rapidly onto a quasi-steady manifold where inhibitory fluctuations are \emph{enslaved} to excitatory ones. Making the timescales explicit:
\begin{equation}
\tau_I \frac{\partial}{\partial t} \delta r_I(k,t) \approx 0
\quad \implies \quad 
\delta r_I(k,t) \approx \alpha(k) \, \delta r_E(k,t),
\end{equation}
with
\begin{equation}
\alpha(k) \approx \frac{\tau_I s_I \, \widetilde w_{IE} \, \widehat{W}_{IE}(k)}{1 - \tau_I s_I \, \widetilde w_{II} \, \widehat{W}_{II}(k)}.
\end{equation}
In the homogeneous limit $k=0$, this reduces to
\begin{equation}
\alpha_0 = \frac{\tau_I s_I \, \widetilde w_{IE}}{1 - \tau_I s_I \, \widetilde w_{II}}.
\end{equation}

Substituting this relation into the excitatory dynamics gives the \emph{effective slow-mode equation}:
\begin{equation}
\tau_E \frac{\partial}{\partial t}\delta r_E(k,t) 
= \left[ -1 + s_E \widetilde w_{EE} \, \widehat{W}_{\mathrm{eff}}(k) \right] \delta r_E(k,t),
\end{equation}
where the \emph{effective spatial connectivity} in Fourier space is
\begin{equation}
\widehat{W}_{\mathrm{eff}}(k) = \widehat{W}_{EE}(k) - \frac{\widetilde w_{EI}}{\widetilde w_{EE}} \, \alpha(k) \, \widehat{W}_{EI}(k).
\end{equation}

The \emph{inverse Fourier transform} of $\widehat{W}_{\mathrm{eff}}(k)$ defines the \emph{effective spatial kernel} $W_{\mathrm{eff}}(\mathbf{r})$, which typically exhibits a \emph{Mexican-hat shape}: a central peak of recurrent excitation surrounded by broader inhibition. This structure supports spatially localized activity: $W_{\mathrm{eff}}(0) > 0$ allows local amplification (bursts), while $\int W_{\mathrm{eff}}(r)\,dr \lesssim 0$ ensures global stability. Near the bifurcation, the network effectively resides on a \emph{slow manifold} where
\begin{equation}
r_I(\mathbf{r},t) \simeq \alpha \, r_E(\mathbf{r},t),
\end{equation}
and fluctuations within this subspace produce the spatially structured, transiently sustained activity characteristic of balanced networks near criticality.

\subsubsection{Irregular Homogeneous Firing State}

In the previous sections, we focused on non-stationary, burst-like dynamics emerging at phase transition. Here, we shift attention to the contrasting regime of homogeneous networks, where activity is approximately uniform across space and neurons fire irregularly without forming localized bumps. We compare the synaptic dynamics under this homogeneous regime of firing with burst dynamics in the main text as well as section ~\ref{subsec:Cross-Correlation} in Appendix.

In the homogeneous case, where external input is uniform across space, the recurrent conductances are expressed as:
\begin{equation}
G_{\alpha\beta}^* = w_{\alpha\beta} r_\beta^* \, \mathcal{W}_{\alpha\beta}, \qquad 
\mathcal{W}_{\alpha\beta} = \int W_{\alpha\beta}(x-x') dx',
\end{equation}
where $w_{\alpha\beta}$ is the scalar synaptic weight, $r_\beta^*$ is the stationary firing rate of population $\beta$, and $\mathcal{W}_{\alpha\beta}$ is the total integral of the spatial kernel (normalized such that $\mathcal{W}_{\alpha\beta} = 1$ for probability kernels).

The total mean input to population $\alpha$ (including external excitatory and inhibitory drives) is:
\begin{equation}
I_\alpha^* = \sum_{\beta\in\{E,I\}} G_{\alpha\beta}^* (V_R^\beta - V_\alpha^*) + \sum_{\beta\in\{E,I\}} G_{\alpha\beta}^{\rm ext} (V_R^\beta - V_\alpha^*),
\end{equation}
where $G_{\alpha\beta}^{\rm ext} = w_{\alpha\beta}^{\rm ext} r_{\alpha\beta}^{\rm ext} \tau_{s, \alpha\beta}^{\rm ext}$ represents the external conductance drive.

The \emph{balanced-state condition} requires that recurrent and external currents approximately cancel:
\begin{equation}
\sum_{\beta\in\{E,I\}} \left[ G_{\alpha\beta}^* + G_{\alpha\beta}^{\rm ext} \right] (V_R^\beta - V_\alpha^*) \approx 0.
\end{equation}

Explicitly, for the excitatory population ($V_\alpha^* = V_E^*$):
\begin{equation}
w_{EE} \mathcal{W}_{EE} r_E^* (V_R^E - V_E^*) + w_{EI} \mathcal{W}_{EI} r_I^* (V_R^I - V_E^*) + \sum_{\beta} G_{E\beta}^{\rm ext} (V_R^\beta - V_E^*) \approx 0,
\end{equation}
and similarly for the inhibitory population. The stationary rates $r_E^*$ and $r_I^*$ are therefore linearly related. Solving for $r_I^*$ as a function of $r_E^*$ in the excitatory balance equation:
\begin{equation}
r_I^* \approx - \frac{w_{EE} \mathcal{W}_{EE} (V_R^E - V_E^*) r_E^* + \sum_{\beta} G_{E\beta}^{\rm ext} (V_R^\beta - V_E^*)}{w_{EI} \mathcal{W}_{EI} (V_R^I - V_E^*)}.
\end{equation}

By adjusting these parameters, the network is tuned to a fluctuation-driven balanced state, remaining subthreshold and stable while allowing for irregular spiking. 

For spatially structured external input, the balance condition is evaluated in Fourier space. Denoting the Fourier transforms of kernels by $\widehat{W}_{\alpha\beta}(k)$, the mode-$k$ balance reads:
\begin{equation}
\sum_\beta \left[ w_{\alpha\beta} \widehat{W}_{\alpha\beta}(k) \widehat{r}_\beta(k) + \widehat{G}_{\alpha\beta}^{\rm ext}(k) \right] (V_R^\beta - V_\alpha^*) \approx 0.
\end{equation}

Linearizing around the balanced state, the effective mode-dependent gain is:
\begin{equation}
\mathcal{G}_\alpha(k) = \frac{1}{g_{\rm tot}^\alpha} \sum_\beta s_\beta \Big[ w_{\alpha\beta} \widehat{W}_{\alpha\beta}(k) r_\beta^* + G_{\alpha\beta}^{\rm ext} \widehat{W}_{\beta}^{\rm ext}(k) \Big],
\end{equation}
with stability requiring $|\mathcal{G}_\alpha(k)| < 1$ for all $k$. Broad recurrent kernels ($\sigma_{\alpha\beta}$ large) suppress high-$k$ fluctuations, stabilizing the homogeneous state, whereas narrow external inputs can selectively excite intermediate modes, potentially destabilizing local balance.

\subsection{Scale-Free Bursts in Random and Strictly Local Networks}
\label{subsubsec:StochasticDynamics}

\begin{figure}[hbtp]

    \centering

    \includegraphics[width=1\linewidth]{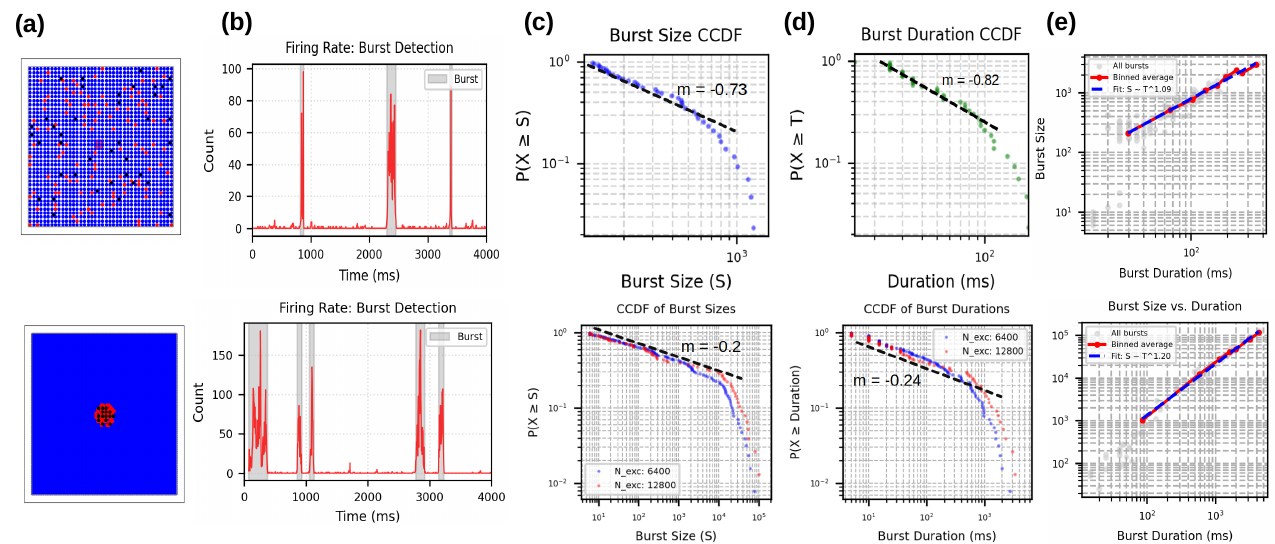}

    \caption{\textbf{Scale-Free Bursts in Random and Strictly Local Networks.} a) Schematic of postsynaptic excitatory (red) and inhibitory (black) neurons connected to central excitatory neurons. Top: Random network; Bottom: Locally connected network. b) Burst detection methodology: events are defined by network activity exceeding a threshold determined by the background firing rate under external input. c) Complementary Cumulative Distribution Function (CCDF) of burst sizes with a fitted power law. d) CCDF of burst durations with a fitted power law. e) Average burst size vs. duration (lifetime) with a fitted power-law scaling relationship.}

    \label{fig:avalnches_net}

\end{figure}

In our model, we observe scale-free behavior and critical scaling near a subcritical Hopf bifurcation. In this regime, stochastic fluctuations induce intermittent transitions between a quiescent state and a balanced active state, which may take the form of a high-activity fixed point or a limit cycle. As the system approaches the non-equilibrium critical point ($\Lambda^* \to 0$), the distributions of burst sizes $S$ and durations $\tau$ follow power laws,
\begin{equation}
P(S) \propto S^{-\tau_s}, \qquad P(\tau) \propto \tau^{-\alpha_t}.
\end{equation}
The critical exponents in the oscillatory regime depend on the network
dimensionality. Figure~\ref{fig:avalnches_net} illustrates avalanche dynamics
in both random and locally connected networks. Our recurrent network
typically operates in the under-damped neighborhood of the Bogdanov-Takens
(BT) point, where the lead eigenvalues acquire an imaginary component. For
comparison, in the vicinity of a pure saddle-node bifurcation
($\Im[\lambda]=0$), the return dynamics are monotonic and effectively
one-dimensional. In this limit, the network reduces to a critical branching
process belonging to the mean-field directed percolation universality class,
characterized by the universal exponents $\tau_s = 1.5$ and $\alpha_t = 2.0$
\cite{ehsani2023scale,ehsani2022self}. This transition from monotonic to
oscillatory return induces a unique universality class where the exponents
shift significantly from the standard 2D Directed Percolation (2D-DP)
(Table~\ref{tab:exponents}). This indicates that critical dynamics at the
edge of oscillation/synchronization place the system outside the standard
directed percolation universality class \cite{munoz2018colloquium}. In
oscillatory criticality, the scaling exponent $\gamma \approx 1.1$--$1.2$ is
markedly lower than in 2D directed percolation ($\gamma \approx 1.67$),
indicating an approximately linear relationship between size and duration,
which is characteristic of stochastic hovering around a limit cycle or
fixed-point solution. E--I coupling limits burst volume through inhibitory
feedback, while spiral dynamics stretch their duration, producing temporally
long but volumetrically thin bursts. This regime preserves scale-free
flexibility and supports rhythmic transients.

A candidate mechanism is provided by \cite{liang2020hopf}, who show that in
mean-field excitation--inhibition networks tuned to a Hopf bifurcation, the
squared oscillation amplitude $X=r^2$ obeys a first-passage-time problem with
demographic (shot) noise, $dX/dt = aX - X^3 + h + \sqrt{X}\,\xi(t)$, with $a$
tuned to the marginal point $a=0$ and $h$ a small effective drive. Because the
noise scales as $\sqrt{X}$, this maps onto an exactly solvable Bessel-type
diffusion whose absorbing-boundary statistics give a continuously tunable
duration exponent $\alpha_t = 3/2 - 2h$, rather than a fixed
universality-class value -- the same near-marginality mechanism underlying
the crackling-noise relation used elsewhere in this work. The authors note
this applies to marginal bifurcations generally, not uniquely to the Hopf
case.

Their derivation is for a \emph{supercritical} Hopf normal form, whereas our
network operates near a Bogdanov--Takens point, where the relevant line is a
\emph{subcritical} Hopf bifurcation terminating on a saddle-node of limit
cycles. The corresponding quintic-stabilized normal form, $dr/dt = ar + br^3
- r^5$ ($b>0$), gives $dX/dt = 2aX + 2bX^2 - 2X^3$. For excursions well below
the unstable limit-cycle amplitude, the cubic and quintic terms are
subleading and short avalanches should inherit the same relation, $\alpha_t
\to 3/2 - 2h$ as $a\to0^-$. Longer avalanches are instead governed by the
nearby saddle-node fold where the stable and unstable limit-cycle branches
annihilate  and are expected to cross over to a second regime with
$\alpha_t \to 2$ for generic (uniform) reinjection, or to be exponentially
suppressed if the network is not tuned close enough to the fold.

The $3/2$ asymptote is not specific to the Bessel construction: a purely
diffusive escape with no deterministic drift, $dr = \sqrt{2D}\,dW$, gives the
classical inverse-Gaussian first-passage density
\begin{equation}
f_T(T) = \frac{R}{\sqrt{4\pi D}}\,T^{-3/2}\exp\!\left(-\frac{R^2}{4DT}\right),
\label{eq:fpt_diffusion}
\end{equation}
with the same $T^{-3/2}$ tail. Two structurally different constructions
converging on the same exponent suggests $3/2$ is a robust lower anchor for
near-marginal escape, not an artifact of one model.

A separate relation connects duration and size exponents directly. If burst
size $S=\int A\,dT$ for an envelope $A\propto T^{\kappa}$, then $S\propto
T^{1+\kappa}$ and $f_T(T)\propto T^{-\alpha_t}$ maps onto $f_S(S)\propto
S^{-\tau_s}$ with
\begin{equation}
\tau_s = \frac{\alpha_t+\kappa}{1+\kappa}.
\label{eq:size_duration}
\end{equation}
For a roughly constant envelope ($\kappa\approx0$, consistent with Fig.~3a),
$\tau_s\approx\alpha_t$. Against Table~\ref{tab:exponents} this holds only
approximately: $\tau_s\approx1.82$ vs.\ measured $1.73$ (Rand-MF),
$\approx1.24$ vs.\ $1.2$ (2D-Latt), and $\approx1.7$ vs.\ $1.55$ (Local+LR) --
close enough to suggest a near-constant envelope, not close enough to replace
a direct $\tau_s$ measurement.

This two-regime picture is broadly consistent with all three networks in
Table~\ref{tab:exponents}. Liang et al.'s empirical relation,
$\alpha_t\approx1.21\tau_s-0.154$, predicts $\alpha_t\approx1.72$ (Local+LR,
measured $1.7$), $\approx1.94$ (Rand-MF, measured $1.82$), and $\approx1.30$
(2D-Latt, measured $1.24$) -- all in the right neighborhood without being
exact, as expected if the relation captures part but not all of the
mechanism. All three measured values also fall within the $[3/2,2]$ interval
spanned by our two derived asymptotes, which is consistent with but does not
confirm the two-regime picture; confirming it would require resolving the
shape of the duration distribution itself. The low $\gamma\approx1.1$--$1.2$
shared by all three networks, against $\gamma=2.0$ for mean-field DP and
$1.67$ for 2D-DP, argues against a simple DP interpretation and is broadly in
line with other bistable/near-synchronization ("hybrid-type") transitions
reported to show anomalously low $\gamma$ without a settled analytical
explanation \cite{buendia2021hybrid}. We present the subcritical extension
above as one candidate mechanism, not a confirmed one.

\begin{table}[h!]
\centering
\begin{tabular}{@{}llccccc@{}}
\toprule
\textbf{Exponent} & \textbf{Symbol} & \textbf{MF-Br} & \textbf{2D-DP} & \textbf{Local+LR} & \textbf{Rand-MF} & \textbf{2D-Latt} \\ \midrule
Size Dist.     & $\tau_s$   & $1.50$ & $1.27$ & $\approx 1.55$ & $\approx 1.73$ & $\approx 1.2$  \\
Duration Dist. & $\alpha_t$ & $2.00$ & $1.45$ & $\approx 1.70$ & $\approx 1.82$ & $\approx 1.24$ \\
Scaling Exp.   & $\gamma$   & $2.00$ & $1.67$ & $\approx 1.23$ & $\approx 1.10$ & $\approx 1.2$  \\ \bottomrule
\end{tabular}
\caption{Comparison of critical exponents in the oscillatory critical model
against the directed percolation universality class. MF-Br: mean-field
branching process (DP universality class). 2D-DP: two-dimensional directed
percolation. Local+LR: local 2D excitation with sparse long-range
connections, the network analyzed in the main text (Fig.~3, System No.~1,
near the Bogdanov--Takens point). Rand-MF: purely random, sparse network with
mean-field-like connectivity, analyzed in this appendix. 2D-Latt: purely
local 2D lattice network, analyzed in this appendix.}
\label{tab:exponents}
\end{table}

\subsection{Cross-Correlation and Synaptic Weight Changes in Different Dynamic States}
\label{subsec:Cross-Correlation}

\subsubsection*{Input Covariance and Plasticity in Irregular Homogeneous and Avalanche Regimes}

Consider a network of $N$ neurons distributed in one- or two-dimensional space, with neuron $i$ at $\mathbf{x}_i$. Connectivity decays with distance, e.g.,
\[
w_{ij} = w(|\mathbf{x}_i - \mathbf{x}_j|) = w_0 \exp\big(-|\mathbf{x}_i - \mathbf{x}_j|/\lambda\big),
\]
where $\lambda$ sets the spatial range. The network exhibits two regimes: asynchronous irregular spiking and bursting (``avalanche'') activity with power-law distributed burst sizes.

For each neuron $i$, spike times $\{t_n^{(i)}\}$ and synaptic inputs are recorded. The total input is
\[
I_i^\mathrm{tot}(t) = I_i^\mathrm{exc}(t) - I_i^\mathrm{inh}(t),
\]
with summed excitatory $I_i^\mathrm{exc}(t)$ and inhibitory $I_i^\mathrm{inh}(t)$ contributions. Pairwise covariances of total input,
\[
\mathrm{Cov}\big(I_i^\mathrm{tot}, I_j^\mathrm{tot}\big) = \langle I_i^\mathrm{tot}(t) I_j^\mathrm{tot}(t) \rangle - \langle I_i^\mathrm{tot}(t) \rangle \langle I_j^\mathrm{tot}(t) \rangle,
\]
are computed as a function of inter-neuron distance $r_{ij} = |\mathbf{x}_i - \mathbf{x}_j|$, giving a spatial covariance function.

\begin{figure}[hbtp] \centering \includegraphics[width=1\linewidth]{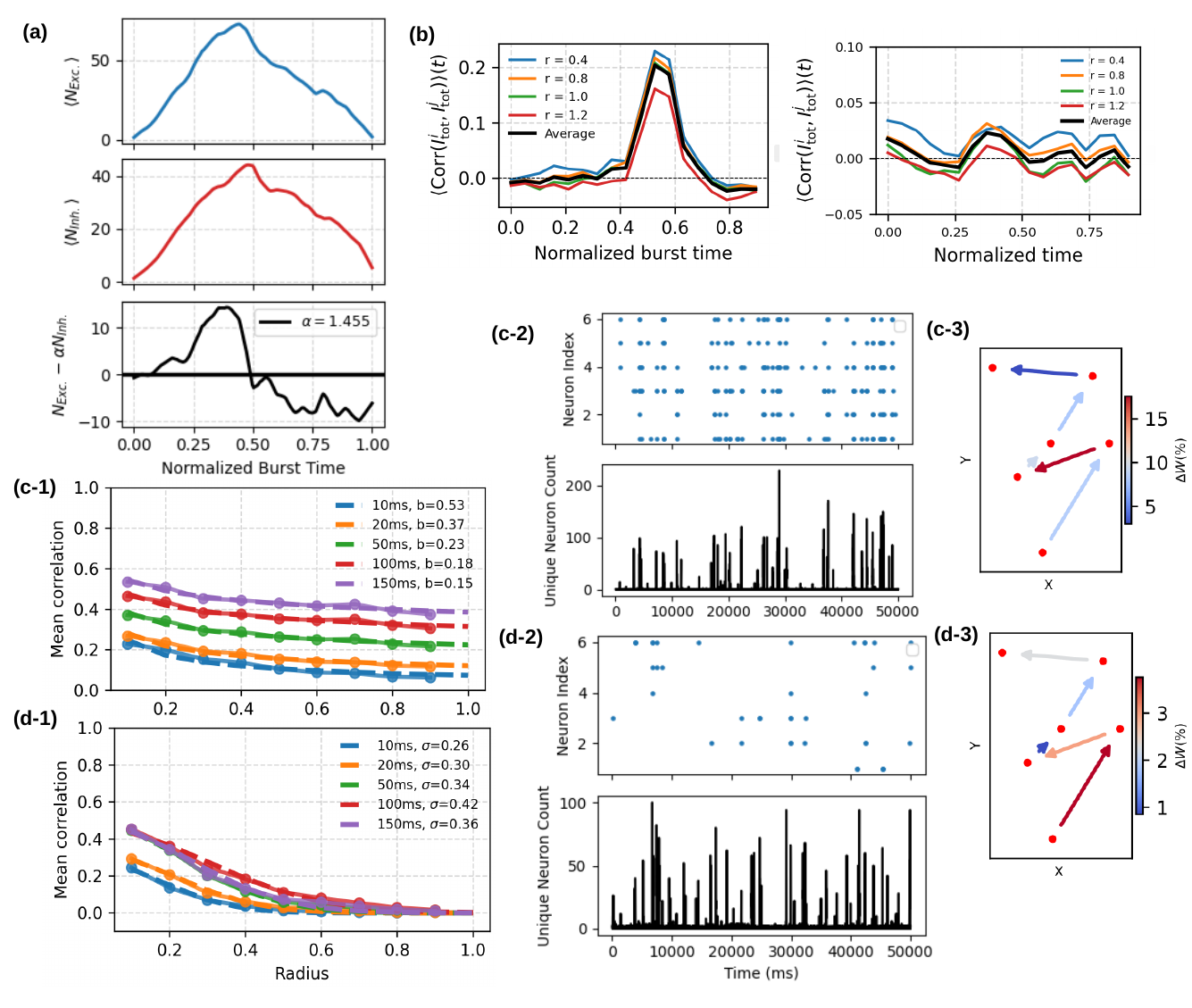} \caption{\textbf{Total input and firing rate covariances between neurons in distinct dynamical regimes} (a) Temporal profiles of average excitation, inhibition, and their difference across all bursts (normalized by burst duration). (b) (Left) Cross-correlations between neurons total input at various spatial distances, measured at different time points within the normalized burst. (Right) same when network is set at homogeneous firing state, normalized by total simulation time. (c) System in a critical regime with scale-free avalanches: (c-1) Spatial firing rate correlations of neurons as a function of distance, calculated across various time bin sizes with power-law fits. (c-2) Raster plot of neurons firing along a chain (nodes distributed randomly over the spatial domain) alongside the total population activity. (c-3) Synaptic changes via STDP along the chain after 50 s of simulation. (d) Same as (c) but for a system showing non-critical, locally confined bursts: (d-1) The spatial correlation function decays exponentially. (d-2) Raster plot of neuron chains shows less synchronous activation across the chain. (d-3) Synaptic changes via STDP are weaker due to the absence of long-range correlations.} \label{fig:corr_chain} 
\end{figure}

Near criticality, inhibition dynamically \emph{shadows} excitation with a short delay $\delta$ (Fig.~\ref{fig:corr_chain}a) :
\[
I_i^\mathrm{inh}(t) \approx I_i^\mathrm{exc}(t - \delta),
\]
generating temporally asymmetric covariances. This produces a positive peak in the cross-covariance for short lags, reflecting transient synchronization across neurons (Fig.~\ref{fig:corr_chain}b-Left). Covariances are strongest for nearby neurons but extend over longer distances, enabling burst propagation and local STDP potentiation while global E/I balance prevents runaway activity. Here, bursts are defined as contiguous activity periods and interpolated to a normalized time axis $t_\mathrm{norm}\in[0,1]$. Phase-resolved covariance traces between central and surrounding neurons, averaged across bursts, reveal how input synchrony evolves. Peaks in Fig.~\ref{fig:corr_chain}b-Left indicate high-synchrony phases, while troughs correspond to desynchronized periods. Comparing the two regimes highlights long-range covariances in the avalanche regime versus weak, rapidly decaying covariances in the asynchronous state.

In the asynchronous regime, excitation and inhibition nearly cancel, producing weak residual fluctuations. Covariances of total input and firing rates are small across all distances (Fig.~\ref{fig:corr_chain}b-Right). Symmetric STDP produces minimal weight change in this regime as shown in the Fig.~\ref{fig:line_attractor} of the main text.  

\subsubsection*{Cross Corelation in Spatial Domain: Spontaneous Replay of Spatially Disjoint Sequences in a Balanced Network}

In a spatially extended excitatory--inhibitory (E--I) network, the spatial structure of neuronal activity can be characterized by correlations between neurons, either in their spike trains or in the total synaptic input they receive. These correlations reveal how collective activity propagates across the network and how it is shaped by network interactions.

At the \emph{critical point}, corresponding to the onset of bursting or avalanche-like activity, spatial correlations decay slowly with distance and are often well described by a power law,
\begin{equation}
C(r) \sim r^{-\kappa},
\end{equation}
reflecting long-range coordination of neuronal activity (Fig.~\ref{fig:corr_chain}-c). The correlation length becomes large, indicating that fluctuations are dominated by collective network dynamics rather than by local connectivity alone. In the middle and right panels of Fig.~\ref{fig:corr_chain}-c, one can see that increasing the spatial extent of the interaction kernel further enhances these correlations, effectively promoting coordinated activity across spatially disjoint neuronal assemblies.

In contrast, in the \emph{asynchronous regime} away from the critical point, correlations decay rapidly with distance, typically approximately exponentially,
\begin{equation}
C_{\mathrm{spk}}(r;L) \sim e^{-r/\xi},
\end{equation}
as shown in Fig.~\ref{fig:corr_chain}-d. Here, activity is largely uncorrelated beyond the local neighborhood, and the network exhibits only short-range coherence.

\subsubsection*{Multiple Sequence Replay}

\begin{figure}[hbtp]
    \centering
    \includegraphics[width=1\linewidth]{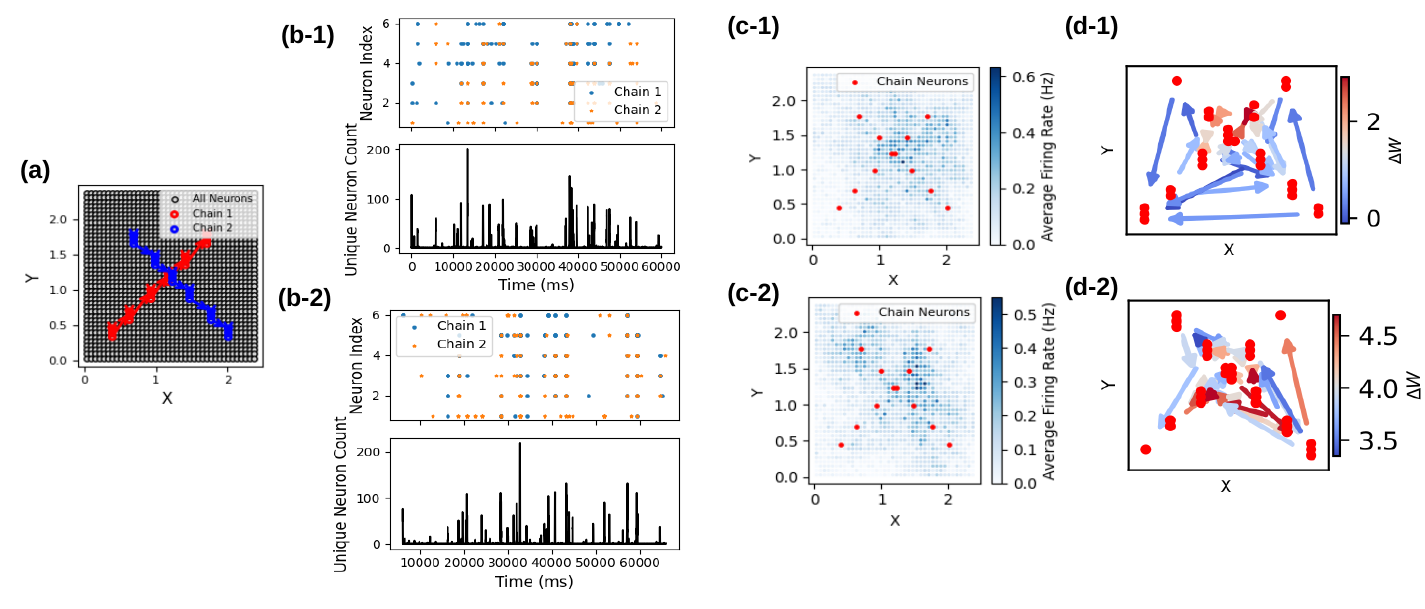}
    \caption{\textbf{Interaction of two neuronal sequences in scale-free and locally confined bursting regimes.} (a) Embedding of the neural sequences through active exploration of two straight paths. (b-1) Raster plot of sequence replay and population firing rate in the network with locally confined bursts. (b-2) Same as (b1) for the network with scale-free bursts.(c-1 and c-2) Firing rate maps of neurons during spontaneous activity for the two systems. (d-1) Synaptic weight changes between neurons for existing connections in the network with locally confined bursts. (d-2) Same for the network with scale-free bursts.   }
    \label{fig:two_seq}
\end{figure}

In Fig.~\ref{fig:two_seq}, we show a simple network composed of two sequences sharing a common node. These sequences are embedded with a spatial spacing larger than the width of the recurrent connectivity kernel. We compare the effects of synaptic changes in a system showing localized bursts (Fig.~\ref{fig:two_seq}b1) versus one with scale-free avalanches (Fig.~\ref{fig:two_seq}b2). Sequence replay, expressed through bursts of varying sizes, can potentiate the synapses between the two sparsely linked chains. These synaptic modifications are significantly more effective in the critical avalanche regime due to the presence of long-range correlations.
In theory, this could lead to the emergence of new subsequences that support task-relevant information processing or the construction of a 2D map from one-dimensional sequences. However, this mechanism requires further careful investigation.

\subsection{Effect of Different Spontaneous Firing Regimes on Behavioral Performance}
\label{subsec:Regimes and Behaviour}

\begin{figure}[hbtp]
    \centering
    \includegraphics[width=1\linewidth]{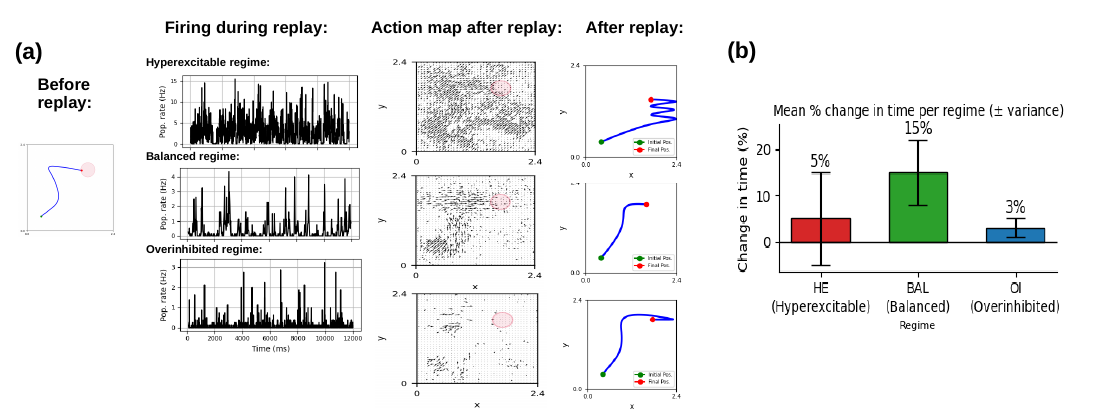}
    \caption{\textbf{Effect of the spontaneous firing regime on behavioral performance.}
    (a) Left: a learned trajectory after exploration and before the replay period.
    Middle: spontaneous firing rates in the three regimes (hyperexcitable, balanced
    and overinhibited). Right: the effect of replay on the action encoded at each
    spatial location, and the resulting behavioral trajectory after replay.
    (b) Mean percentage change in trial time after replay, computed over 15 different
    non-optimal trajectories, for the hyperexcitable (HE), balanced (BAL) and
    overinhibited (OI) regimes. Error bars show the variance across trajectories.}
    \label{fig:regime_behaviour}
\end{figure}

To test how the spontaneous firing regime affects the behavioral benefit of replay,
we started from 15 non-optimal trajectories learned during exploration and measured
the change in trial time after a replay period in each of three regimes
(Fig.~\ref{fig:regime_behaviour}a). The balanced regime produced the largest
improvement, reducing trial time by 15\% on average (Fig.~\ref{fig:regime_behaviour}b).
Both imbalanced regimes led to much smaller improvements, but for different reasons.

In the hyperexcitable regime, weak inhibition leads to widespread over-activation
during replay. Many place cells are co-active at once, so plasticity strengthens
associations between locations and actions that are not causally related. These
excessive, incorrect associations overwrite parts of the useful policy, resulting
in only a small mean improvement (5\%) and a high variability across trajectories
(variance 10) due to the spurious associations.

In the overinhibited regime, strong inhibition confines excitation to local
neighbourhoods. Replay events cannot propagate across the environment, so no
long-range associations between distant locations are formed and the policy is
updated only locally. This yielded the smallest improvement (3\%) and the lowest
variability (variance 2): the behavior changes consistently, but very little.

Together, these results indicate that a balanced spontaneous firing regime is needed
for replay to improve behavior. It provides enough excitation for replay to link
distant locations, and enough inhibition to keep the resulting associations selective.

\subsection{ Transient Diffusion in 1D Balanced Network }
\label{subsec:Transient Diffusion}

In this section, we study spontaneous activity in response to spatially and temporally homogeneous external noise in a 1D network consisting of 20 E–I populations, each containing 2,000 excitatory and 500 inhibitory neurons, arranged in a ring. 

\begin{figure}[hbtp]
    \centering
    \includegraphics[width=1\linewidth]{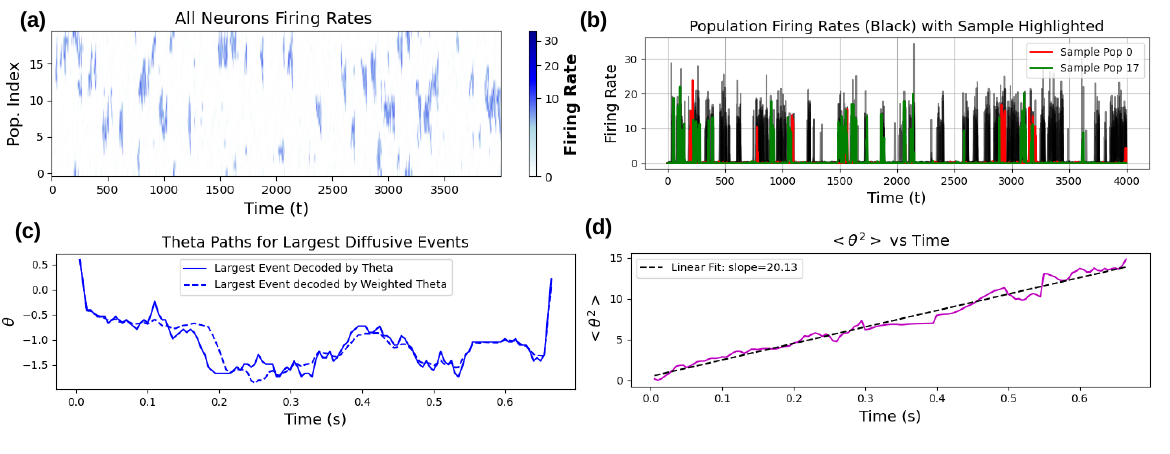}
    \caption{\textbf{Transient diffusion in an E--I network on a ring.} 
    (a) Firing rates of 20 excitatory populations over time. 
    (b) Overall average network firing rate and individual firing rates for selected populations. 
    (c) Decoded trajectory of the phase $\theta$ for the largest burst event, calculated using weighted and unweighted averages of population activity. 
    (d) Mean squared displacement $\langle \theta^2(t) \rangle$ as a function of time, characterizing the diffusive (random walk) nature of the decoded phase. }
    \label{fig:diffusion_1D}
\end{figure}

As we observed for the 2D spatial network, near a global bifurcation in a balanced E–I network, the homogeneous system exhibits transient, burst-like activity rather than stable bumps (Fig.~\ref{fig:diffusion_1D}a--b). Small inhomogeneities in connectivity or input can nucleate localized bursts that are metastable: they form at one location, are rapidly suppressed by inhibitory feedback, and reappear elsewhere. This results in fast, effectively diffusive propagation of bursts across the network. By decoding the network activity in time bins, one can observe a random-walk-like behavior in 1D ($\Theta$) during these variable-size bursts (Fig.~\ref{fig:diffusion_1D}c--d).

In networks with broad excitatory kernels and near-neutral translation modes, stable bumps persist longer and diffuse slowly:
\[ \qquad D_{\rm slow} \sim \frac{\sigma_{\rm noise}^2}{\gamma_{\rm pin}} \ll D_{\rm fast}.
\]
Here $D_{\rm slow}$ is set by external noise and pinning stiffness $\gamma_{\rm pin}$, while $D_{\rm fast}$ depends primarily on burst lifetime and spatial heterogeneity. This explains why near-critical balanced networks exhibit rapid, metastable propagation of transient bursts, contrasting with slow, noise-driven wandering of stable bumps.

\end{document}